\documentclass[aoas,preprint]{imsart}

\RequirePackage[OT1]{fontenc}
\RequirePackage{amsthm,amsmath}
\RequirePackage[]{natbib}
\usepackage{booktabs} 
\usepackage{bm}
\usepackage{graphicx} 
\usepackage{pdfpages}

\usepackage{xr}
\arxiv{arXiv:1810.00121}

\startlocaldefs
\numberwithin{equation}{section}
\theoremstyle{plain}

\endlocaldefs

\begin{document}

\begin{frontmatter}
\title{Discovering Interactions Using Covariate Informed Random Partition Models}
\runtitle{}

\begin{aug}
\author{\fnms{Garritt L. Page} \snm{}\thanksref{m1, m2}\ead[label=e1]{page@stat.byu.edu}},
\author{\fnms{Fernando A. Quintana} \snm{}\thanksref{m3, m4}\ead[label=e2]{quintana@mat.uc.cl}}
\and
\author{\fnms{Gary L. Rosner} \snm{}\thanksref{m5}
\ead[label=e3]{ grosner1@jhmi.edu}
}

\runauthor{Page et al.}

\affiliation{Brigham Young University \thanksmark{m1} and BCAM - Basque Center for Applied Mathematics \thanksmark{m2} and
		Pontificia Universidad Cat\'olica de Chile \thanksmark{m3} and
Millennium Nucleus Center for the Discovery of Structures in Complex Data
\thanksmark{m4} and John Hopkins University \thanksmark{m5}}

\address{Department of Statistics, \\Provo, Utah\\
Departamento de Estad\'istica, \\Santiago, Chile\\
\printead{e1}\\
\phantom{E-mail:\ }\printead*{e2}}

\address{Department of Biostatistics, \\Baltimore, Maryland \\
\printead{e3}\\
}
\end{aug}

\begin{abstract}
Combination chemotherapy treatment regimens created for patients diagnosed
with childhood acute lymphoblastic leukemia have had great success in
improving cure rates. Unfortunately, patients prescribed these types of
treatment regimens have displayed susceptibility to the onset of
osteonecrosis. Some have suggested that this is due to pharmacokinetic
interaction between two agents in the treatment regimen (asparaginase and
dexamethasone) and other physiological variables. Determining which
physiological variables to consider when searching for interactions in
scenarios like these, minus {\it a priori} guidance, has proved to be a
challenging problem, particularly if interactions influence the response
distribution in ways beyond shifts in expectation or dispersion only. In
this paper we propose an exploratory technique that is able to discover
associations between covariates and  responses  in a general way. The
procedure connects covariates to responses flexibly through dependent
random partition distributions, and then employs machine learning
techniques to highlight potential associations found in each cluster. We
provide a simulation study to show utility and apply the method to data
produced from a study dedicated to learning which physiological predictors
influence severity of osteonecrosis multiplicatively.
\end{abstract}



\end{frontmatter}

\section{Introduction}\label{sect:intro}

In studies that collect covariate measurements on subjects/experimental
units in addition to a response,  it is of principal interest 
to determine which covariates influence the response and in what way. This seemingly
benign statistical problem has been seriously considered for many years
with methods ranging from exploratory techniques to more model-based
procedures.  Perhaps the reason why research dedicated to this problem
persists is that in practice building a statistical model is as much an
art as a science. This is a consequence of important or ``significant''
associations depending completely on the factors included in the model.
Therefore, knowing {which} effects to include in a model is crucial, but
this information is rarely known {\it a priori}.  Since this information
is rarely known, an approach that is commonly used in practice proposes
fitting a saturated model (a model containing all possible covariates or
predictors) and then employing some type of multiplicity test correction,
model selection criterion, shrinkage method, or stochastic search to locate
``significant'' factors \cite[see, e.g.,][]{Tibshirani:1996,
smith&kohn:1996, george&mcculloch:1997, ishwaran&rao:2005,
chung&dunson:2009, scott&berger:2010,  mitra&mueller&ji:2017}. These
methods have been shown to work well in many instances
(\citealt{gomezetal:2017}). However, when multiplicative effects and/or
nonlinear associations are present and are of interest, a fairly common
scenario \citep[see, e.g.][]{LogLinearModelsForGeneAssociation,LimHastie},
the process of identifying associations becomes much more problematic as
the saturated model can quickly become unwieldy. This happens to be the
case in the study under consideration, where we examine which factors
affect the severity of osteonecrosis in children treated for acute
lymphoblastic leukemia (ALL). Interest lies in learning how time-varying
physiological and other baseline covariates such as gender, triglycerides,
and body mass index (BMI) influence disease severity in an additive or
multiplicative fashion. Employing the saturated model approach here with
all possible two-way and/or three-way interactions  becomes
computationally expensive and inferentially difficult, so knowing which
covariates to consider is important.

Adding to the complexity of the scenario just described, methods used to
identify associations (either additive or multiplicative) are geared
towards discovering covariates that only influence the mean of the
response distribution. It is completely plausible that interactions also
influence the variance or even the shape of the response distribution, and
most methods are not equipped to detect these types of associations.
Regression methods that allow covariates to influence the entire response
(or error) distribution have been developed and are commonly known as
density regression \citep[see][]{fan&qiwei&tong:1996,
BayesianDensityRegression}. There has been work in density regression that
simultaneously carries out variable selection
(\citealt{tokdar&zhu&ghosh:2010, shen&ghosal:2016}), but they require that
all effects of interest be included in a model. Though it may be possible
to extend work done in density regression to incorporate multiplicative
effects, {\it a priori} information would be necessary to guide which of
these effects to consider. In light of this, an exploratory procedure that
is able to discover possible associations in a general way, prior to model
fitting, would be appealing.

There is a growing literature dedicated to methods developed with the main
purpose of explicitly identifying interactions without restricting
themselves to variable selection.
\cite{reich&kalendra&storlie&bondell&fuentes:2012} develop a statistical
emulator and devise a procedure that allows them to learn how the inputs
of a stochastic computer model influence the output (which is a
distribution). \cite{bien2013} use the Lasso,  and \cite{lim&hastie:2015}
the group Lasso to detect interactions from hierarchically coherent models
(i.e., two-way interactions are present only if both main effects are
present). \citealt{kapelner&bleich:2016} employ Bayesian additive
regression trees (BART) to identify interactions while \cite{linero}
develop methods based on Bayesian decision tree ensembles that
incorporate an additive component. 
XGBoost~\citep{chen&geustringXgboost:2016} is a scalable machine learning
system for tree boosting, that has also been employed to discover
interactions. In more recent work \cite{ferrari&dunson:2019} use a factor
model to induce models that include interactions and \cite{agrawal19a} search for interactions in high dimensions.


The methods just cited are developed for specific scenarios while what
we propose is  more general and can be employed with any data model,
making the  procedure essentially ``model free.''  This is important for
our application as the response is ordinal, rendering many of the works
just cited unavailable.  Thus, the exploratory procedure we propose
discovers associations where anything is fair game in the sense that
associations could be additive and/or multiplicative and could influence
any aspect of the response density (i.e., mean, spread, shape, etc.)
regardless of the data model that will eventually be employed.  Our goal
is ambitious and admittedly being able to discover all possible
interactions in the general way we are proposing is presumably not
possible.  Thus, we do not claim that the exploratory approach detailed in
the sequel discovers all ``significant'' interactions; rather it provides
guidance to practitioners by highlighting possible interactions with very
little overhead.

We finish our literature review noting that there is a growing literature
dedicated to so-called subgroup analysis, i.e., the study of heterogenous
treatment effects among subgroups of a study population. Subgroups are
typically identified or defined based on specific values in the covariate
space.  Since subpopulation treatment effects are of principal interest,
focus is placed on studying the interaction between a covariate and a
treatment (see e.g., \citealt{simon, berger&wang&shen, varadhan&wang:2014,
schnell&tang&offen&carlin, liu&sivaganesan&laud&muller:2017,
su&pena&liu&levine:2018, henderson&louis&rosner&varadhan:2020}). Our approach is more general in nature but can
be used as an exploratory tool to help discover which covariates interact
with the treatment and provide guidance by spotting subpopulations of
potential interest.

As a point of terminology, in what follows we use the term ``interaction''
to denote something more general than what is referred to in a linear
model setting.  Here, an interaction exists if the response distribution
is in some way influenced by specific combinations of covariate values.
This general conception includes the special case where all the influence
is carried by only one of the covariates. Thus, we recommend carefully investigating the ``interactions'' detected by our method to gain insight into how they (and the covariates they contain) influence the response distribution. This highlights the exploratory nature of our proposal.

The remainder of the article is organized as follows. In
Section~\ref{sect:data} we detail the data collected from the study that
motivated this work. Section~\ref{sect:background} provides an overview of
our approach with some background on dependent random partition distributions
and association rules. Section~\ref{sect:InterDiscProc} provides more
details of our approach and Section~\ref{sim.studies} describes two
simulation studies. In Section~\ref{InteractionsInStudy} we
apply the procedure to the osteonecrosis data and Section~\ref{sect:disc}
contains a brief discussion.

\section{Osteonecrosis Study}\label{sect:data}

The study we consider was designed to learn more about factors affecting
risk for osteonecrosis in children suffering from acute lymphoblastic
leukemia (ALL). With combination chemotherapeutic regimens, five-year
survival is around 85\% overall for childhood ALL, with some subgroups'
rates well above 90\% (\citealt{ALL}).  These regimens include the drug
asparaginase and the steroid dexamethasone. Some have suggested that there
is a pharmacokinetic interaction between these two agents, leading to
greater inter-patient variability and severe adverse events
(\citealt{kawediaetal:2011}). The principal aim of the motivating study
was to learn about relationships between physiological characteristics and
treatment characteristics, and how these relationships influence
susceptibility to osteonecrosis as a result of therapy. The analysis
considers a number of physiological covariates measured on each patient,
including low-density lipoprotein (LDL),  high-density lipoprotein (HDL)
levels, body mass index (BMI), and others.  In addition to these
covariates, the data include plasma levels of dexamethasone, cortisol, and
asparaginase at various times during each patient's course of treatment
(baseline, week 7, week 8) to explore how the pharmacokinetics (PK) of
these substances influenced the risk and severity of osteonecrosis, and
whether other factors interact with the drugs' PK.  Finally, demographic
variables include age at diagnosis, gender, and race. In total 23
predictors were considered with numerical summaries provided in Tables
\ref{RaceGender} and \ref{CovariateSummary}. Out of the 400 patients in
the study, we have complete data vectors for 234, and we focus on these.

Table \ref{RaceGender} displays the gender-by-race distribution with each gender being
equally represented. A third binary covariate (called {LowRisk}) which
indicates the prognosis of the patient's leukemia (low or high risk of
relapse) is also included. This variable should be highly influential
because exact treatment regimens are based on patients' risk factors
relating to prognosis of their leukemia.  Of the 234 subjects,  109 were
in the low-risk group while 125 were in the high risk. Table \ref{CovariateSummary} provides a numerical summary of continuous covariates. Notice that a few of them are
highly right skewed.  Empirical correlations between time varying
covariates suggest that temporal dependence is also present. This dependence
can be accommodated in our approach and is formally considered in the
model developed by \cite{barcella&deiorio&favaro&rosner:2018}. We briefly
comment that week 12 measurements were also collected, but very
irregularly.  This resulted in many incomplete covariate vectors, and for
this reason we only consider measurements taken up to week 8.

The response measured in this study reflects the severity of osteonecrosis
ranging on an ordinal scale from 0 for none up to 4 for high grade. Table
\ref{OsteonecrosisGrade} contains the number of patients by grade of
osteonecrosis. We expect, but do not force, the covariates' influences on
osteonecrosis grade to be multiplicative, but it is not clear which
covariates to pair together when exploring multiplicative effects. This is
something we hope to discover.

\begin{table}[htp]
\caption{Number of patients in each gender by race category.}
\begin{center}
\begin{tabular}{l | c c c c c} \toprule
Gender & Asian & Black & Hispanic & Other & White\\ \midrule
Female & 1 & 21 & 8 & 7 & 96\\ \midrule
Male & 2 & 19 & 8 & 3 & 69 \\ \bottomrule
\end{tabular}
\end{center}
\label{RaceGender}
\end{table}%

\begin{table}[htp]
\caption{Numerical summary of the nineteen continuous covariates measured on each subject in the osteonecrosis study.}
\begin{center}
\begin{tabular}{l | r r r r r r} \toprule
Covariate &   Min. & Q1 &  Q2  &  Mean  & Q3  &  Max. \\ \midrule
AgeAtDiagnosis 			& 1.02 & 3.20 & 5.23 & 6.77 & 9.68 & 18.84 \\
  DexClWk07 				& 1.99 & 9.03 & 14.69 & 21.83 & 25.03 & 458.07 \\
  DexClWk08 				& 1.62 & 8.09 & 11.67 & 14.07 & 16.62 & 61.71 \\
  CortisolBaseline 			& 0.25 & 4.07 & 5.67 & 7.46 & 8.22 & 59.41 \\
  CortisolWk07 				& 0.11 & 4.93 & 7.12 & 9.00 & 10.34 & 55.16 \\
  CortisolWk08 				& 0.00 & 0.32 & 0.58 & 1.11 & 1.01 & 30.27 \\
  HDLBaseline 				& 11.30 & 39.05 & 48.60 & 56.28 & 59.25 & 271.00 \\
  HDLWk07 				& 10.70 & 41.92 & 52.50 & 59.10 & 67.20 & 198.00 \\
  HDLWk08 				& 1.90 & 29.45 & 44.00 & 50.31 & 61.50 & 238.00 \\
  LDLBaseline 				& 13.00 & 62.00 & 82.00 & 83.95 & 104.75 & 195.00 \\
  LDLWk07 				& 15.00 & 57.25 & 80.00 & 82.04 & 100.38 & 218.00 \\
  LDLWk08 				& 5.00 & 50.25 & 68.00 & 73.44 & 91.50 & 179.00 \\
  TriglyceridesBaseline 		& 20.00 & 58.62 & 87.50 & 104.60 & 124.75 & 472.00 \\
  TriglyceridesWk07 			& 20.00 & 55.02 & 74.50 & 114.79 & 132.50 & 885.00 \\
  TriglyceridesWk08 			& 15.90 & 75.25 & 122.50 & 220.75 & 286.00 & 1029.50 \\
  AlbuminBaseline 			& 3.30 & 4.00 & 4.20 & 4.17 & 4.40 & 5.10 \\
  AlbuminWk07 				& 2.10 & 3.20 & 3.90 & 3.75 & 4.30 & 4.80 \\
  AlbuminWk08 				& 2.20 & 3.10 & 3.60 & 3.56 & 4.00 & 4.80 \\
  BMIBasline				& 12.19 & 15.68 & 16.77 & 18.05 & 19.23 & 38.48 \\
  AsparaginaseAntibodyAUC 	& 0.42 & 1.07 & 3.54 & 14.20 & 24.83 & 93.85 \\  \bottomrule
\end{tabular}
\end{center}
\label{CovariateSummary}
\end{table}%

\begin{table}[htp]
\caption{Grade of osteonecrosis and the number of patients diagnosed for
each one.}
\begin{center}
\begin{tabular}{l | c c c c c} \toprule
Osteonecrosis Grade & 0 & 1 & 2 & 3 & 4\\ \midrule
Number of Subjects & 73 & 115 & 27 & 17 & 2 \\ \bottomrule
\end{tabular}
\end{center}
\label{OsteonecrosisGrade}
\end{table}%

\section{Background and Preliminaries}\label{sect:background}

The exploratory procedure we develop consists of three stages. In this
section we  introduce each one and then provide notation and background
information necessary to make ideas concrete.

The first step consists of connecting covariates to the response by way of
a dependent random partition model. These types of models create
partitions of the data taking into account homogeneity among covariates,
and in particular, some of these models, such as the PPMx
by~\cite{PPMxMullerQuintanaRosner}, aim specifically at constructing priors
for which two individuals with similar covariate values are more likely to
cocluster. The resulting partition is, {\it a posteriori}, weighted by the
corresponding cluster sizes, the likelihood component and the similarity
of covariates that belong to a cluster. This feature has proven to be
quite useful in capturing many types of nonlinearities present in the
data, including combined changes in shape, location and scale of the
predictive distribution as the predictors change. Since these types of
models are closely related to discrete random probability models of the
type used in Bayesian nonparametrics, which are well known for their
flexibility~\citep[see, e.g.][]{mueller&quintana&jara&hanson}, in theory
any statistical model that produces (explicitly or implicitly) a covariate
dependent partition may be employed. This includes the dependent Dirichlet
process mixture (DDP) model of \cite{MacEachernddp} and its variants
(e.g., ANOVA DDP of \citealt{deiorio&mueller&rosner&maceachern:2004}). As
explained below, we adopt here the PPMx model.

In the second step the estimated partition from step 1 is used to identify
potentially interesting interactions. One possible way of employing the
partition to pinpoint potential interactions is to identify covariates
that seem to influence cluster formation. This would imply that an
association exists among the influential covariates and the response (a
necessity for an interaction to exist). Identifying influential covariates
can be done by  determining which covariates have the same (or similar)
values for many individuals in a cluster. An unsupervised learning
technique developed to carry out this type of search establishes so called
association rules~\citep[Chapter 6]{DataMiningBook}. There is precedence
to using  association rules as a tool to identify interactions (see
\citealt{wseas2014advances}), but we use them to rank potential
interactions based on their ``importance.'' Ranking  interactions based on
their ``importance'' is, implicitly, something all procedures described in
Section~\ref{sect:intro} do. Many of the procedures, however, provide very
little guidance or criteria on how to determine which interactions are
simply noise. This motivates the inclusion of a third step to our
exploratory approach.

In the third step of our procedure  we verify that a detected interaction
indeed affects the response distribution.  This is done by comparing
posterior predictive densities that are based on specific values from
covariates that were identified by an association rule.  If the posterior
predictive densities do not change as a function of the covariates found
in the association rule, then their association with the response is weak
at best. (As a side note, \cite{gabry&simpson&vehtari&betancourt&gelman}
advocate using a posterior predictive distribution as an exploratory or
confirmatory tool.) We now provide more pertinent background information
for dependent random partition models and association rules.

\subsection{Dependent Random Partition Models} \label{drpm}

We begin by introducing some  notation.  Let  $i = 1, \ldots, m$ index the
$m$ experimental units in a designed experiment or $m$ subjects in an
observational study. Further, let $\rho_m = \{S_1, \ldots, S_{k_m}\}$
denote a partitioning (or clustering) of the $m$ units into $k_m$ subsets
such that $i \in S_j$ implies that unit $i$ belongs to cluster $j$.  A
common alternative notation that specifies a partitioning of the $m$ units
into $k_m$ clusters is to introduce $m$ cluster labels $s_1, \ldots, s_m$
such that $s_i = j$ implies $i \in S_j$. We will use $Y_i$ to denote the
$i$th subject's response variable with $\bm{Y} = (Y_1, \ldots, Y_m)$
denoting an $m$ dimensional response vector and $\bm{Y}^{\star}_j =  \{Y_i
: i \in S_j\}$  the $j$th cluster's response vector. Similarly, let
$\bm{X} = (X_{1}, \ldots, X_{m})$ denote a covariate vector and
$\bm{X}^{\star}_j = \{X_i:i \in S_j\}$  a partitioned covariate vector.
When $p$ covariates are measured on each individual, both continuous and
qualitative, then $\bm{X}_i = (X_{i1}, \ldots, X_{ip})$ will denote the
$i$th individual's $p$-dimensional covariate vector and with a slight
abuse of notation set $\bm{X}^{\star}_j = (\bm{X}^{\star}_{j1}, \ldots,
\bm{X}^{\star}_{jp})$ where $\bm{X}^{\star}_{jh} = \{X_{ih}:i \in S_j\}$
for $h= 1,\ldots,p$. Thus depending on context, $\bm{X}^{\star}_j$ could
possibly be a super vector of stacked patients' covariate vectors.

We now introduce notation associated with the model that will be employed to connect $\bm{X}$ to $\bm{Y}$ through $\rho_m$.
A dependent random partition prior distribution is assigned to $\rho_m$.
This prior distribution parametrized by $\bm{\eta}$, will be denoted using
$RPM_X(\bm{\eta})$. Once a prior for $\rho_m$ is specified,  we will make
use of $f(\bm{Y} \mid  \rho) = \prod_{j=1}^{k_n} f_j(\bm{Y}^{\star}_j)$ as a
data model where $f_j(\bm{Y}^{\star}_j) = \int \prod_{i \in S_j} f_j(Y_i
\mid \bm{\theta}^{\star}_j)dG_0(\bm{\theta}^{\star}_j)$,  $f_j(\cdot \mid 
\bm{\theta}^{\star}_j)$ denotes the likelihood for $\bm{Y}^{\star}_j$, and $G_0$ a
prior for cluster specific parameters $\bm{\theta}^{\star}_j$. Note that
unit-specific parameters can be connected to their cluster-specific
counterpart via $\bm{\theta}_i = \bm{\theta}^{\star}_{s_i}$. Alternatively, the data
model can be written hierarchically  using the cluster labels $s_1,
\ldots, s_n$ in the following way
\begin{align}\label{model2}
Y_i \mid \bm{\theta}^{\star}, s_i & \stackrel{ind}{\sim} f_{s_i}(\bm{\theta}^{\star}_{s_i}), \ \mbox{for} \ i = 1, \dots, m \nonumber \\
\bm{\theta}^{\star}_{\ell} & \stackrel{iid}{\sim} G_0, \ \mbox{for} \ \ell = 1, \ldots, k_m, \\
\rho_m\mid \bm{X} & \sim RPM_X(\bm{\eta}). \nonumber
\end{align}

Notice that covariates do not appear in the data model so that  neither
pre-specified associations nor their forms are required. As a result,
$\bm{Y}$ is only connected  to $\bm{X}$ through the posterior distribution
of $\rho_m$ (denoted by  $\pi(\rho_m\mid \bm{Y},\bm{X}))$ which will
facilitate our interaction search. There are a number of computational
techniques that have been developed to fit model \eqref{model2}. Most are
some variant of MCMC that depends on the exact specification of
$RPM_X(\bm{\eta})$. We opt to employ algorithm 8 of
\cite{MCMCSamplingMethodsForDPmixtureModels} when fitting model
\eqref{model2}, since it is a general algorithm that can be used in a
variety of settings.

As already discussed, there are many possibilities for $RPM_X(\bm{\eta})$.
In what follows we employ the covariate dependent random partition model
(PPMx) described in \cite{PPMxMullerQuintanaRosner} (and further explored
in \citealt{page&quintana:2018}) because of its flexibility in being able
to easily incorporate all types of covariates. The exact form of
$RPM_X(\bm{\eta})$ when adopting the PPMx is
\begin{align*}
RPM_X(\bm{\eta}) \propto \prod_{j=1}^{k_m}
c(S_j)g(\bm{X}^{\star}_j\mid \bm{\eta}),
\end{align*}
where $c(S)$ is a set function that measures the chance that elements in $S$
co-cluster {\it a priori}, and $g(\bm{X}^{\star}\mid \bm{\eta}) = \int
\prod_{i \in S_j} q(\bm{X}_i\mid \bm{\xi}^{\star})
q(\bm{\xi}^{\star}\mid \bm{\eta})d\bm{\xi}^{\star}$ is a similarity function
that produces higher values for $\bm{X}^{\star}$'s that contain covariate
values that are more similar. Here, $q(\cdot\mid\cdot)$ denotes an
auxiliary probability model (likelihood and prior) that has no effect in
the data model and is used only to introduce the desired effect in the
similarity function.  When multiple covariates of different types are
available \cite{PPMxMullerQuintanaRosner} suggest using the following
multiplicative form
\begin{align}\label{ProductSimilarityFunction}
g(\bm{X}^{\star}_j\mid \bm{\eta}) = \prod_{\ell = 1}^{p} g(\bm{X}^{\star}_{j \ell}\mid \bm{\eta}).
\end{align}

\subsection{Association Rules}

Association rules are used to discover patterns or relations among a large
collection of variables. They are typically denoted using $\{A
\}\Rightarrow \{B\}$ where $A$ and $B$ define a subset of the covariate
space that  does not share any variables. Connecting the idea to the
osteonecrosis study, a possible association rule is  $\{A  =
\mbox{AgeAtDiagnosis}  \in [3,5)\} \Rightarrow \{B  = \mbox{DexClWk07} \in
[9, 12)\}$ which would indicate that if  $AgeAtDiagnosis$ is between 3 and
5, then $DexClWk07$ is between 9 and 12.

There are two criteria to evaluate the importance of $\{A \}\Rightarrow
\{B\}$. The first, called {\em support}, is the proportion of patients
whose covariate vector contains both $A$ and $B$.  The second criterion,
called {\em confidence}, is the fraction of patients that display $B$
among those that display $A$.  Our criteria to determine which covariate
pairs to consider when looking for interactions will be those with large
support and confidence. For a more detailed overview of association rules
we refer interested readers to \citet[chapter 14]{HTF} or \citet[chapter
6]{DataMiningBook}.

The typical notation used for association rules highlights the fact that
they are by nature directional.  However,  our only purpose in using
association rules is to identify pairs of covariates that are associated
with each other and possibly influence the distribution of $Y$ in a
non-additive way. Because of this, we will use $X_1 \Leftrightarrow X_2$
to denote that at least one association rule contained the pair $X_1, X_2$
in some fashion as the direction of the association is immaterial for our
purposes.

\section{Interaction Discovery Procedure}\label{sect:InterDiscProc}

In this section we provide more details of our three-step interaction
discovery procedure,  after which we provide two simulation studies. In
what follows we will refer to the three-step procedure as the Random
partition, Association rule, Interaction Discovery procedure (or the  RAID
procedure).

A natural way of employing $\pi(\rho_m \mid  \bm{Y}, \bm{X})$ in the second
stage of the exploratory procedure is to produce a point estimate of
$\rho_m$ using methods found in \cite{dahl:2020} (or some other
alternative) and then apply association rules to each of the resulting
$\bm{X}^{\star}_j$s. However, as noted by \cite{wade:2017}, there might be
substantial variability associated with the partition point estimate.
Therefore, we instead apply association rules to the resulting
$\bm{X}^{\star}_j$s for each (or a subset) of the MCMC draws collected.
This approach provides a means of propagating uncertainty associated with
$\rho_m$ through the exploratory procedure.  Since the interactions
identified will be based on specific clusters, they are local in the same
way that treatment effects are local in subset analysis. That is, the
interaction may not remain consistent across the entire population. (We
explore this in the simulation study of  Section~\ref{sim.study.2}.)  We
briefly note that applying association rules to each of the
$\bm{X}_j^{\star}$'s is crucial to finding interactions as they are
connected to $\bm{Y}$ through $\pi(\rho_m \mid  \bm{Y}, \bm{X})$.  Applying
association rules directly to $\bm{X}$ would make interaction detection
impossible as there is no connection to $\bm{Y}$.

As stated previously, the third step consists of using posterior
predictive densities to confirm interactions. To make this step concrete,
consider an example with three binary covariates. Let $p_{ijk}(Y_0\mid \bm{Y},
\bm{X})$ denote the posterior predictive density for $X_1 = i$, $X_2 = j$,
and $X_3 = k$ with $i,j,k \in \{0,1\}$ and assume that within a cluster,
the association rule with highest total support and confidence is $X_1
\Leftrightarrow X_2$.  To verify that there does indeed exist what we call
an interaction between $(X_1, X_2)$, we test the following hypothesis
(after fixing $X_3=k$ to its empirical median)
\begin{align}\label{BNP2}
 H_0: p_{00k}(Y_0\mid \bm{Y}) = p_{01k}(Y_0\mid \bm{Y}) = p_{10k}(Y_0\mid \bm{Y}) = p_{11k}(Y_0\mid \bm{Y}).
 \end{align}
If the hypothesis is rejected, then we conclude that $X_1$ and $X_2$
interact.  That is, the predictive distribution of $Y$ is in some way
influenced by ${X_1}$ and/or ${X}_2$, and the specific way in which this
occurs can be later explored separately.  A similar approach would be employed if association rules highlighted any of the other possible two-way interactions (i.e., $X_1
\Leftrightarrow X_3$, $X_2 \Leftrightarrow X_3$).  We also consider the possibility of a three-way interaction in Section \ref{InteractionsInStudy}.

There are a number of procedures
that might be selected to test the hypothesis found in \eqref{BNP2}. The
P\'olya tree procedure outlined in \cite{chen&hanson} is quite flexible
and powerful, and thus we opt to use it in what follows. However, any
other procedure that is able to formally test the hypothesis in
\eqref{BNP2} is completely valid. Just as with Chen and Hanson's (2014) method,
most methods employed to test \eqref{BNP2} will produce a ``$p$-value.''
We  use the ``$p$-values'' as a validation tool and thus, there is no
guarantee that Type 1 error rates are preserved. That said, the simulation
in the Section \ref{sim.study.2} suggests that error rates are not far
from the size of the test.

\section{Simulation Studies} \label{sim.studies}
In this section we detail two simulation studies carried out to explore
the RAID procedure's ability to detecting interactions.  The first is
based on a toy example that is included to provide insight into each of
the three steps of our approach.  The second is a more realistic setting
that is based on the covariate structure found in the osteonecrosis data
set.  In both simulation studies a continuous response variable is used
even though the response in the osteonecrosis study is ordinal.  This was
done so that we could compare our approach to other procedures found in
the literature. As far as we are aware, no other procedure is able to
accommodate an ordinal response.  This highlights the ``model-free''
property of our approach.

In order to connect $\bm{Y}$ to $\bm{X}$
through $\pi(\rho_m\mid \bm{Y}, \bm{X})$,  we fit the following particular
case of model \eqref{model2}  to each generated synthetic data set in both simulation studies
\begin{align}\label{modelNormal}
Y_i \mid \bm{\mu}^{\star}, \bm{\sigma}^{\star}, s_i & \stackrel{ind}{\sim} N(\mu^{\star}_{s_i}, \sigma^{\star}_{s_i}), \ \mbox{for} \ i = 1, \dots, m \nonumber \\
(\mu^{\star}_{\ell}, \sigma^{\star}_{\ell}) & \stackrel{iid}{\sim} N(\mu_0, \sigma^2_0) \times Unif(0, A), \ \mbox{for} \ \ell = 1, \ldots, k_m, \\
\rho_m\mid \bm{X} & \sim RPM_X(\bm{\eta}), \nonumber
\end{align}
with $\mu_0 \sim N(0, 10^2)$,  $\sigma_0 \sim Unif(0, 10)$, and $A=1$. For
the $RPM_X(\bm{\eta})$ we used the PPMx with cohesion $c(S) = M\times(|S|
- 1)!$ connecting the partition model to that induced by a Dirichlet
process (DP) mixture and the ``rich get richer'' property (i.e., a model
that favors a small number of large clusters). $M$ has connections with
the DP dispersion parameter and we used $M=1$.  For the similarity we
employed the auxiliary similarity function (see
\citealt{PPMxMullerQuintanaRosner, page&quintana:2018}). As a result, for
qualitative variables, $q(\cdot \mid  \bm{\xi}^{\star})$ and
$q(\bm{\xi}^{\star}\mid \bm{\eta})$  correspond to multinomial and Dirichlet
density functions, respectively  and for continuous variables they take on
a Gaussian and Gaussian-Inverse-Gamma density functions. Thus, for
qualitative covariates $\bm{\eta}$ is a vector of Dirichlet shape
parameters that we set to 0.1 and for continuous covariates, $\bm{\eta}$
corresponds to the $m_0=$ center, $k_0=$ scale, $\nu_0=$ shape, and
$\kappa_0=$ rate parameters of Gaussian-Inverse-Gamma distribution. After
standardizing continuous covariates, we employed $m_0=0$, $k_0=0.5$,
$\nu_0=1$, $\kappa_0=2$. In this model, the prior parameters that seem to
have the most influence over $\pi(\rho_m\mid \bm{Y}, \bm{X})$ are $A$ and
$k_0$. These parameters determine how much weight is placed on $\bm{Y}$ or
$\bm{X}$ when forming clusters. We explore sensitivity to the
specification of these prior parameters in Section \ref{priorSensitivity}.

\subsection{Simulation Study: Toy Example} \label{sim.study.toyexample}

In this simulation, data sets consisting of three binary covariates are
generated (i.e., $X_i \in \{0,1\}$, for $i = 1,2,3$). The response
distribution $f(Y\mid  X_1, X_2,X_3)$ is made to explicitly depend on $\bm{X}
= (X_1, X_2, X_3)$ in the three ways that are listed in Table
\ref{Response_Distribution_SS}. Each row of the table represents a
possible covariate combination and the columns correspond to a particular
response distribution. Note first from the table that only $X_1$ and $X_2$ influence the
distribution of $Y$, and they do so multiplicatively (i.e., they interact
in the linear model sense). As a control, we consider the case where  the
distribution of $Y$ does not depend on $\bm{X}$ in any way,  which is
denoted by $f_0(Y\mid \bm{X})$. From Table \ref{Response_Distribution_SS} note
that $f_0(Y\mid \bm{X})$ corresponds to a standard normal regardless of the
values of $\bm{X}$. For $f_1(Y\mid \bm{X})$ the variance and shape remain the
same regardless of the value of $\bm{X}$, but the mean changes based on
the values of $X_1$ and $X_2$. For $f_2(Y\mid \bm{X})$, the mean and shape
remain the same, but the variance changes depending on the values of $X_1$
and $X_2$. Lastly, for $f_3(Y\mid \bm{X})$ the mean and variance do not change
but the values $X_1$ and $X_2$ influence the shape. The scenario under
$f_1(Y\mid \bm{X})$ follows the traditional definition of an interaction and
thus can be detected using any number of procedures. The other two
scenarios would not be detected, even if an interaction between $X_1$ and
$X_2$ is explicitly included in a linear model.

\begin{table}[htp]
\caption{List of distributions used in the simulation study. Here $N(a,b)$
denotes a normal distribution with mean $a$ and standard deviation $b$,
$SN(a,b,c)$ denotes a skew-normal distribution with location ($a=10$),
scale ($b=1$), and skew ($c=20$) parameters.  For the two-component
mixture we set $p_1 = p_2 = 0.5$,  $s^2_1 = s^2_2 = 1/16$, and $m_1 = -m_2
=\sqrt{15}/4$. }
\begin{center}
\begin{tabular}{ccc | llll }
\toprule
$X_1$ & $X_2$ & $X_3$ & \multicolumn{1}{c}{$f_0(Y\mid \bm{X})$} & \multicolumn{1}{c}{$f_1(Y\mid \bm{X})$} & \multicolumn{1}{c}{$f_2(Y\mid \bm{X})$} & \multicolumn{1}{c}{$f_3(Y\mid \bm{X})$} \\
\midrule
1 & 1 & 1 & $N(0,1)$ & $N(0,1)$ & $N(0,1)$ & $N(0,1)$\\
1 & 1 & 0 & $N(0,1)$ &  $N(0,1)$ & $N(0,1)$ & $N(0,1)$\\
1 & 0 & 1 & $N(0,1)$ &  $N(2,1)$ & $N(0,3)$ & $SN(a,b,c)$\\
1 & 0 & 0 & $N(0,1)$ &  $N(2,1)$ & $N(0,3)$ & $SN(a,b,c)$\\
0 & 1 & 1 & $N(0,1)$ &  $N(0,1)$ & $N(0,1)$ & $N(0,1)$\\
0 & 1 & 0 & $N(0,1)$ &  $N(0,1)$ & $N(0,1)$ & $N(0,1)$\\
0 & 0 & 1 & $N(0,1)$ &  $N(4,1)$ & $N(0,6)$ & $\sum_{j=1}^2 p_jN(m_j, s^2_j)$\\
0 & 0 & 0 & $N(0,1)$ &  $N(4,1)$ & $N(0,6)$ & $\sum_{j=1}^2 p_jN(m_j, s^2_j)$\\
\bottomrule
\end{tabular}
\end{center}
\label{Response_Distribution_SS}
\end{table}%
For each scenario we generated 1000 data sets each with 500 observations
and fit model \eqref{modelNormal}.  At each MCMC iterate of $\rho_m$,
cluster-specific association rules were gathered using the {\tt apriori}
function found in the {\tt arules} package (\citealt{arules})  of the
statistical software {\tt R}  (\citealt{R-software}).  This function is
based on the {\it apriori} algorithm of
\citealt{Agrawal:1996:FDA:257938.257975}. We considered association rules
from clusters that contained at least 10 subjects/experimental units.
After association rules were gathered using a lower bound support of 0.25 and
confidence of 0.5, we identified the covariate tandem that had the highest
total support and confidence. If this resulted in a tie, then both
covariate tandems were considered. Then for each association rule, the
P\'olya tree-based testing procedure of \cite{chen&hanson} was used to
test the hypothesis in \eqref{BNP2}. The $X$ variable absent from the
association rule was fixed to its empirical median value. The $p$-values
from this testing procedure are derived from permutation tests and values
in Table \ref{HT1} are based on 500 permutations.

Lastly, since the number of posterior predictive draws collected to test
\eqref{BNP2} can be selected somewhat arbitrarily,  the adage
``statistical vs practical significance'' is important here as it would be
undesirable to reject \eqref{BNP2} for differences among predictive
densities that are inconsequential. Thus, when carrying out the test of
\cite{chen&hanson}, we considered $N \in \{ 50, 100, 250\}$ posterior
predictive draws.

\begin{table}[ht]
\caption{$p$-values  averaged over 1000 synthetic datasets  from testing hypothesis 
\eqref{BNP2} using the P\'olya tree method of \cite{chen&hanson}. }
\resizebox{\textwidth}{!}{
\begin{tabular}{l rrrr | rrrr | rrrr |}
\toprule
  & \multicolumn{4}{c}{$N = 50$} & \multicolumn{4}{c}{$N = 100$} & \multicolumn{4}{c}{$N = 250$}  \\  \cmidrule(lr){2-5} \cmidrule(lr){6-9}  \cmidrule(lr){10-13}
  \multicolumn{1}{c}{arules} & $f_0$ & $f_1$ & $f_2$ & $f_3$ & $f_0$ & $f_1$ & $f_2$ & $f_3$  & $f_0$ & $f_1$ & $f_2$ & $f_3$ \\
\midrule
 $X_1$  $\Leftrightarrow$ $X_2$ 	& 0.53 & 0.00 & 0.00 & 0.02 & 0.48 & 0.00 & 0.00 & 0.00 & 0.44 & 0.00 & 0.00 & 0.00  \\
 $X_1$  $\Leftrightarrow$ $X_3$ 	& 0.47 & 0.54 & 0.50 & 0.48 & 0.51 & 0.38 & 0.50 & 0.45 & 0.44 & 0.37 & 0.43 & 0.41  \\
 $X_2$  $\Leftrightarrow$ $X_3$ 	& 0.49 & 0.00 & 0.00 & 0.36 & 0.50 & 0.00 & 0.00 & 0.25 & 0.43 & 0.00 & 0.00 & 0.14  \\
 \bottomrule
\end{tabular}
}
\label{HT1}
\end{table}

In the online supplementary material, we provide a more detailed
description of the simulation study results for each of the 24 possible
association rules, but here we focus on results for testing \eqref{BNP2};
which are found in Table \ref{HT1}. Each column of the table contains the
$p$-values produced by the \cite{chen&hanson} P\'olya tree method
averaged across the 1000 data sets.  The Monte Carlo errors associated with entries found in Table \ref{HT1} are all small with the largest being 0.01.   From Table \ref{HT1} it seems that for
this data generating mechanism, it is sufficient to only consider 50
posterior predictive draws when testing the hypothesis for all three
interaction types.

\begin{figure}[htbp]
\begin	{center}
\vspace{- 1.5cm}
\includegraphics[scale=0.82]{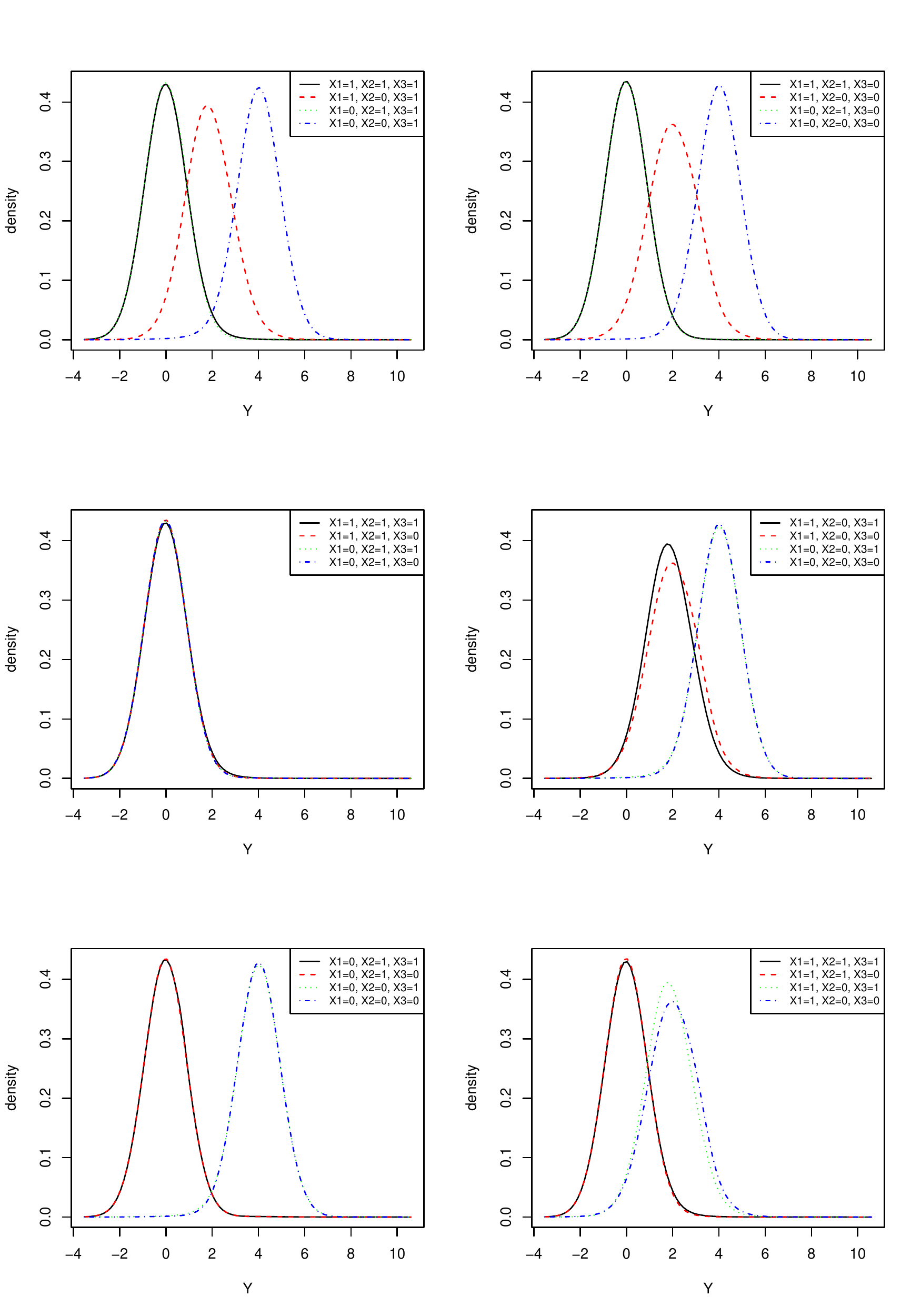}
\vspace{-0.3cm}
\caption{Posterior predictive densities for the six combinations of $X_1$, $X_2$,
and $X_3$ based on a synthetic data set used in the simulation study. }
\label{meanChange}
\end{center}
\end{figure}

From Table \ref{HT1} it also seems that under $f_0$ the average $p$-value is
large in all cases, indicating that the posterior predictive densities in
\eqref{BNP2} are essentially the same regardless of the number of
posterior predictive draws and association rule. This was expected.
Similarly,  $X_1$ $\Leftrightarrow$ $X_3$ under $f_1$, $f_2$, and $f_3$
and $X_2$ $\Leftrightarrow$ $X_3$ under  $f_3$ produced large $p$-values,
which was also expected.  Since there is an interaction between $X_1$ and
$X_2$, the small $p$-values associated with $X_1$  $\Leftrightarrow$ $X_2$
under $f_1$, $f_2$, and $f_3$ were expected. However, notice that under
$f_1$ and $f_2$ the association rule that includes $X_2$ and $X_3$ also
produced small $p$-values for \eqref{BNP2}. This was unexpected. Since
$X_2$ and $X_3$ do not interact, fixing $X_1$ makes it so that
\eqref{BNP2} amounts to testing for the ``marginal effect'' of $X_2$.
Thus, small $p$-values associated with hypothesis \eqref{BNP2} can
indicate an interaction or a marginal effect. To correctly determine
which, further testing can be carried out, or  the posterior predictive
densities being compared in \eqref{BNP2} should be carefully examined.
Here, we carried out the later by providing the posterior predictive densities
in Figure \ref{meanChange}. Estimated densities in the figure correspond
to the first synthetic data set generated based on $f_1$. The first row of
the figure corresponds to densities when $X_3$ is fixed, the second row
when $X_2$ is fixed, and the third when $X_1$ is fixed. Notice that the
densities in the first row are  noticeably different and clearly depend on
the levels of $X_1$ and $X_2$ (i.e., there is an interaction between $X_1$
and $X_2$). Close inspection of rows 2 and 3 of the figure shows that
differences seen in the posterior predictive densities are not due to
$X_3$ at all, but entirely to $X_1$ for the second row and $X_2$ for the
third. Thus, graphs like Figure \ref{meanChange} are able to help
determine if interactions correspond to ``main effects'' or ``interaction
effects'' as they are typically defined in a linear model setting.
Ultimately, the take-home message of Table \ref{HT1} and Figure
\ref{meanChange} is that the exploratory procedure is able to identify
that $X_1$ and $X_2$ interact regardless of the influence that they have
on the distribution of $Y$.

Next we compared the performance of our method to five alternative approaches that 
detect interactions. To this end, to each
generated data set, we applied the following procedures:
\begin{enumerate}
\item linear model with all main effects and two-way interactions;
\item spike \& slab model of \cite{ishwaran&rao:2005}  with all main
    effects and two-way interactions;
\item the hierarchical group-Lasso (hgLASSO) procedure in
    \cite{LimHastie};
\item the BART procedure detailed in \cite{kapelner&bleich:2016};
\item procedure based on XGBoost (\citealt{EIX}).
\end{enumerate}
Each of the listed methods were fit in {\tt R} using the following: {\tt
lm} function for linear model, the {\tt spikeslab} package
(\citealt{ish&raoR-package}) for spike \& slab, the {\tt glinternet}
package (\citealt{glinternetR}) for hgLASSO, the {\tt bartMachine} package
(\citealt{kapelner&bleich:2016}) for BART, and the {\tt xgboost} package
(\citealt{xgboostR}) coupled with the {\tt EIX} package (\citealt{EIX})
for XGBoost. To apply the procedures just listed, some-type of user input
was required to confirm that an interaction was detected. For the sake of
conciseness, we provide specific details of how procedures were
implemented in the online supplementary material.

To study the impact that the RPM prior might have on results, we also used
the RAID procedure coupled with the induced random partition distribution
from the ANOVA DDP. The particular ANOVA DDP we employed is that for which
the atoms were made to depend on  group membership (as defined by the
three categorical covariates), but weights did not. Lastly, we checked the
sensitivity of results to posterior convergence by identifying
interactions based on an MCMC chain that was run for only 500 iterations

The values listed in Table \ref{competitors} are the  proportion of data
sets for which the interaction between $X_1$ and $X_2$ was correctly
identified. The Monte Carlo standard errors associated with proportions in Table \ref{competitors}  were all very small with 0.09 (associated with ``RAID with ANOVA DDP, N = 100'') being the largest.   Notice that when no interaction was present all procedures
performed similarly except for XGBoost and RAID with ANOVA DDP. These two
procedures seemed to be more liberal in their interaction detection.
When the interaction was based on the
expectation of the response distribution (column $f_1$), all procedures detected the interaction except for XGBoost. However, if the
interaction was based on the variance or shape of the response
distribution, the RAID procedure with PPMx was the only one that
consistently detected it. Lastly, the RAID procedure still performed
surprisingly well, even when the number of collected MCMC iterates was
small.

\begin{table}[htp]
\caption{Percent of datasets in the simulation study for which the
interaction between $X_1$ and $X_2$ was correctly identified.}
\begin{center}
\begin{tabular}{l | ccccc} \toprule
\multicolumn{1}{c}{Procedure} & $f_0$ & $f_1$ & $f_2$ & $f_3$ \\ \midrule
Linear Model 					& 0.04 & 1.00 & 0.05 & 0.06 \\
Spike \& Slab 					& 0.08 & 1.00 & 0.05 & 0.09 \\
hgLASSO 					& 0.09 & 1.00 & 0.12 & 0.10 \\
BART 						& 0.00 & 0.99 & 0.00 & 0.00 \\
XGBoost 						& 0.27 & 0.54 & 0.37 & 0.40 \\
RAID with ANOVA DDP, $N=50$ 	& 0.19 & 1.00 & 1.00 & 0.23 \\
RAID with ANOVA DDP, $N=100$ 	& 0.44 & 1.00 & 1.00 & 0.47 \\
RAID with ANOVA DDP, $N=250$ 	& 0.61 & 1.00 & 1.00 & 0.87 \\
RAID with PPMx, $N=50$ 		& 0.06 & 1.00 & 1.00 & 0.96 \\
RAID with PPMx, $N=100$ 		& 0.04 & 1.00 & 1.00 & 1.00 \\
RAID with PPMx, $N=250$ 		& 0.08 & 1.00 & 1.00 & 1.00 \\

\midrule
\multicolumn{5}{c}{Results with only 500 MCMC iterates} \\ \midrule
RAID with PPMx, $N=50$ 		& 0.10 & 1.00 & 1.00 & 0.23 \\
RAID with PPMx, $N=100$ 		& 0.28 & 1.00 & 1.00 & 0.73 \\
RAID with PPMx, $N=250$ 		& 0.13 & 1.00 & 1.00 & 1.00 \\

\bottomrule
\end{tabular}
\end{center}
\label{competitors}
\end{table}%

\subsection{Simulation Study: Osteonecrosis Study Covariate Structure}\label{sim.study.2}

We now describe a simulation study where the number of covariates in each
synthetic data set is similar to that found in the osteonecrosis study.
More specifically, each synthetic data set contained 21 covariates, 19 of
which were continuous and 2 binary. The binary covariate values were
generated using a Bernoulli distribution with probability of success equal
to 0.5. The continuous covariates were generated using a continuous
uniform distribution with support $(-1,1)$. We created interactions that
impact the mean, spread, and shape of the response distribution in two
ways. The first employed the setup established in Table
\ref{Response_Distribution_SS}, except that instead of one noise covariate that was categorical, there are 19
continuous noise covariates.

The second method of generating data sets was based on interactions that
depend on continuous covariates. This was done by using the following data
generating mechanisms to generate response values
\begin{itemize}
\item $Y \sim N(5X_1X_2, \sigma)$ (interaction affecting mean);
\item $Y \sim N(0, \exp\{ 5X_1X_2\}\sigma )$ (interaction affecting
    spread);
\item $Y \sim \pi(X_1, X_2) N(0, \sigma) + (1-\pi(X_1,X_2))\left[
    0.5N(-0.5, \sigma) + 0.5N(0.5, \sigma) \right]$ where $\pi(X_1,
    X_2) = 1/(1 + \exp\{200X_1X_2\})$ (interaction affecting shape).
\end{itemize}
For this setting $X_1$ and $X_2$ were continuous covariates generated
using a uniform distribution with support $(-1,1)$.  Similar to what was
done before, 19 ``noise'' covariates were also included, 17 of which were
continuous (coming from a uniform distribution with support $(-1,1)$) and
two binary (coming from a Bernoulli distribution with probability of
success equal to 0.5).

We explored how the ratio of signal to noise affected the ability to
detect interactions by considering  $\sigma \in \{10^{-3},  10^{-2},
10^{-1},  10^{0}\}$. Finally, we also considered the case when only 60\%
and 25\% of observations exhibited the interaction. This was done by using
a $N(0,1)$ distribution to generate response values for 40\% (or 75\%) of
the units.

To each synthetic data set, the RAID PPMx procedure was fit along with the
six competitors introduced  in
the previous section. For each procedure we recorded if the interaction
was detected (True Positive) and also enumerated the number of falsely
detected interactions (False Positives). We briefly note that since
association rules require discrete variables, before employing association
rules for each cluster, the continuous covariates were dichotomized such
that an equal number of observations belonged to each interval.

Here we only describe results for the case when the interaction was present in 100\%
of the population and include results when the interaction was present in only
60\% or 25\% of the population in the online supplementary material (see
figures S.1-S.4).  Trends were similar in all scenarios, except that it
became more challenging to detect interactions  as the percent of
population that exhibits interaction decreased with the RAID method being
the least impacted when the interaction only applies to a subset of the
population.

\begin{figure}[htbp]
\begin{center}
\includegraphics[width=1.0\textwidth]{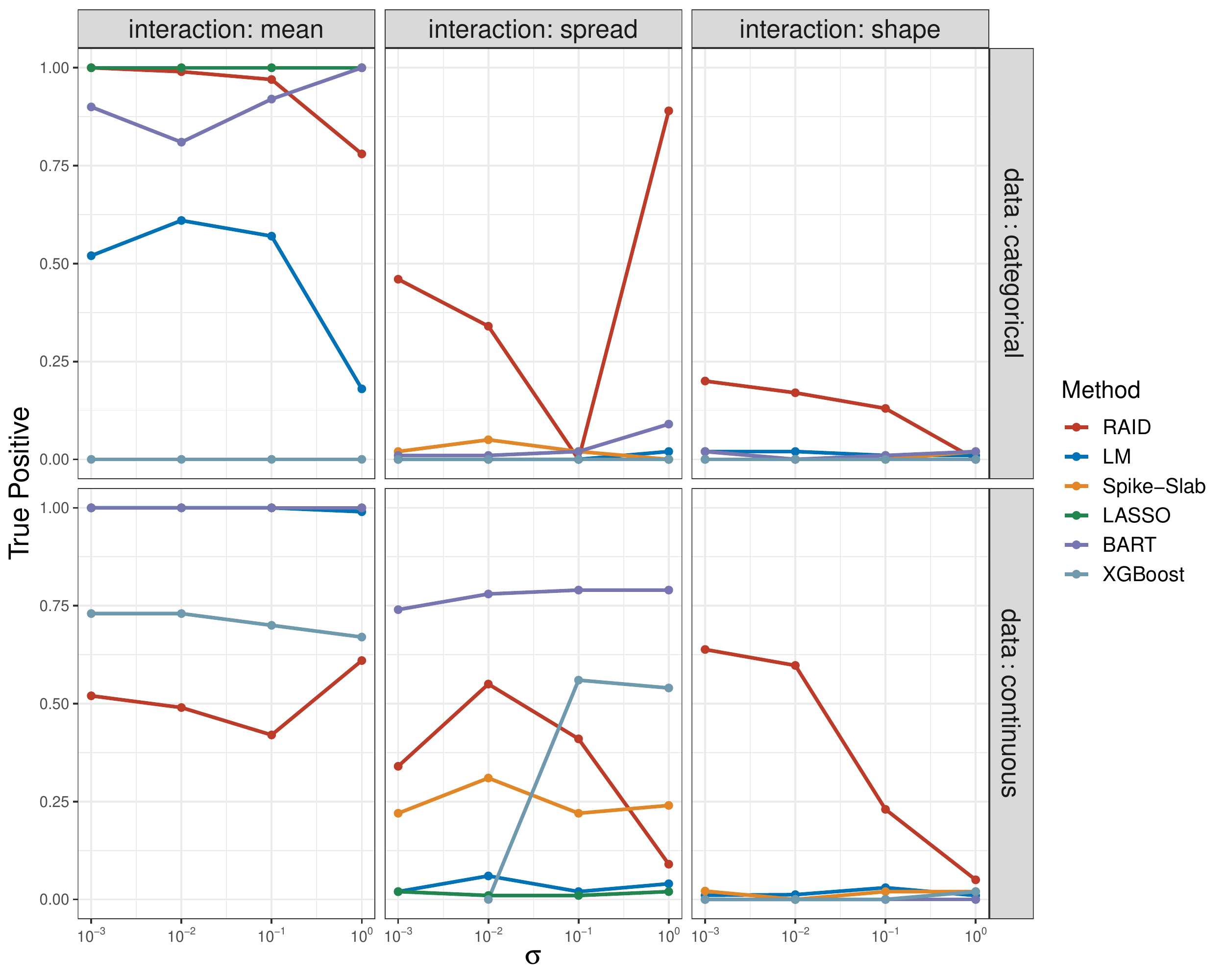}
\caption{Results from the second simulation study: percentage of synthetic
datasets for which the interaction was detected. In each figure, the top
row corresponds to an interaction that is based on categorical covariates
and the bottom row to continuous covariates. } \label{simstud2Results1}
\end{center}
\end{figure}

In Figure \ref{simstud2Results1}, notice that except for XGBoost all
methods did reasonably well in recovering the interaction that affects the
mean between two categorical covariates, with the RAID and LM methods
decaying more when the signal-to-noise ratio decreased.  When interactions
based on continuous covariates affected mean, the RAID method did the
worst (only detecting the interaction in about 50\% of the synthetic
datasets). However, when interactions affected spread or shape, the RAID
procedure is the only method that was able to detect the interaction with
any type of regularity. The BART method won when interaction between
continuous covariates affected the spread and the XGBoost method tended to
perform the worst.

\begin{figure}[htbp]
\begin{center}
\includegraphics[width=1.0\textwidth]{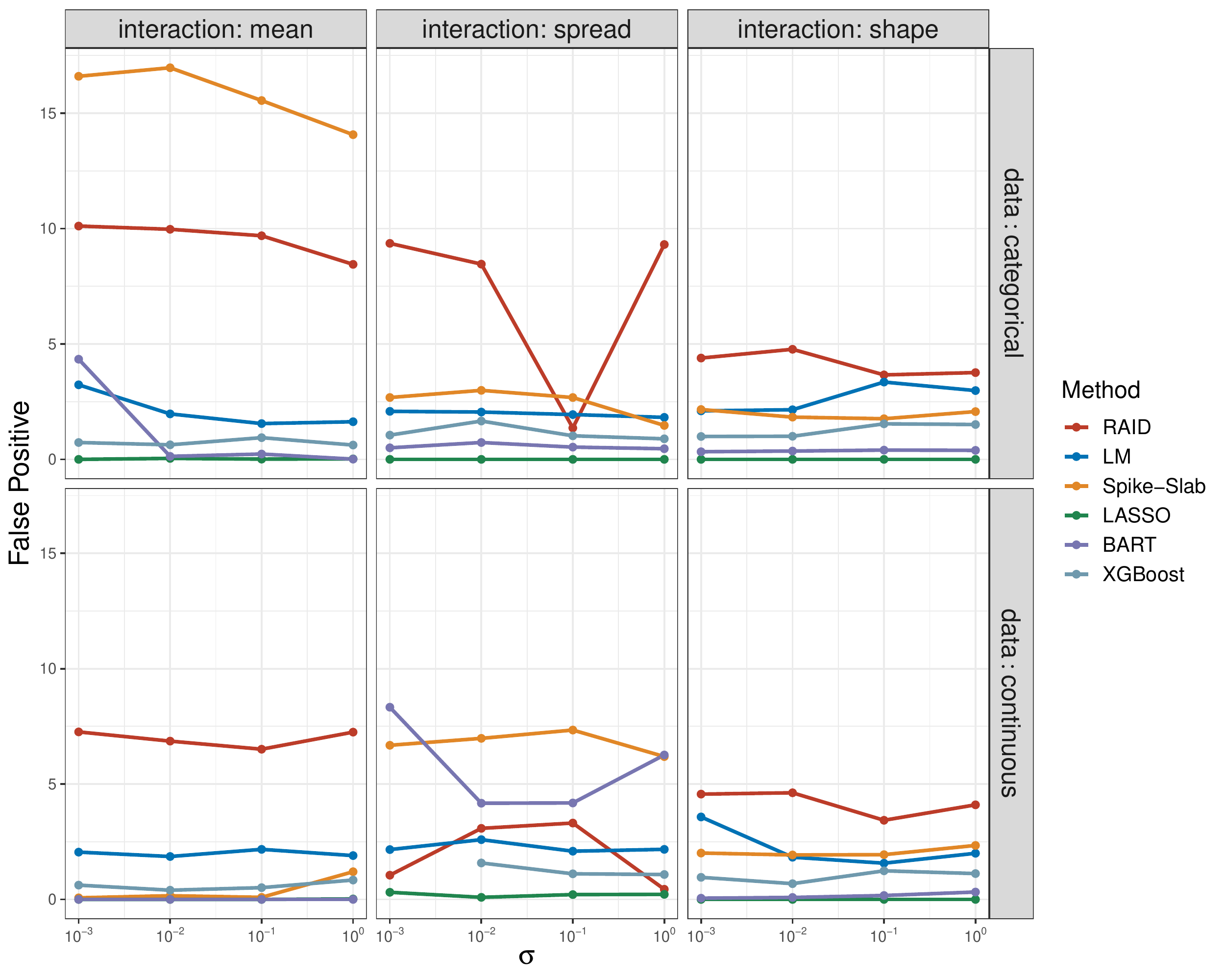}
\caption{Results from the second simulation study: average number of falsely
detected interactions.  In each figure, the top row
corresponds to an interaction that is based on categorical covariates and
the bottom row to continuous covariates. } \label{simstud2Results2}
\end{center}
\end{figure}

Figure \ref{simstud2Results2} displays the false positives. It seems that
the spike \& slab and RAID procedures  tended to produce the most amount
of false positives, while the LASSO and BART method did reasonably well in
avoiding this. Upon further investigation, many of the false positives
detected by the RAID procedure included one of $X_1$ or $X_2$ (covariates
that participated in the interaction). In fact, if interactions that included either $X_1$ or $X_2$ were not classified as false positives, then the number of
false positives detected by the RAID procedure, averaged over all data generating scenarios,
was only 1.01. Thus, as discussed in the previous section, the
RAID procedure detects any combination of covariates that affect the
response distribution in some way.

\section{Interactions in the Osteonecrosis Study} \label{InteractionsInStudy}

We now turn our attention to the osteonecrosis study. Since the response
is an ordinal variable, it is natural to consider latent variable ordinal
data models as those described in  \cite{bao&hanson:2015} and in
\cite{kottas&mulled&quintana:2005}. These types of models are clearly more
complex than model \eqref{model2}, but they permit demonstrating the
flexibility of the exploratory procedure. In fact, any data model can be employed
so long as the response is connected to the covariates through a dependent
random partition model (or analogous predictor dependent clustering
procedures).

For the sake of completeness we detail the ordinal model in its entirety.
Let $Y_i \in \{0,1,2,3,4\}$ be the ordinal response measured on the $i$th
patient $i =1, \ldots, 234$. Further, let $\bm{X}_i = (X_{i1}, \ldots,
X_{i22})$ denote subject $i$'s 22-dimensional covariate vector.  As is
done in \cite{bao&hanson:2015}, real-valued latent scores $Z_1, \ldots,
Z_n$ are introduced such that for $-\infty=\gamma_0 < \gamma_1 < \ldots
\gamma_4 < \gamma_5 = \infty$,
\begin{align*}
Y_i = \ell \Leftrightarrow \gamma_{\ell} < Z_i \le \gamma_{\ell+1} \ \mbox{for $\ell = 0, \ldots, 4$}.
\end{align*}
The appeal of employing the methods described in \cite{bao&hanson:2015}
and \cite{kottas&mulled&quintana:2005} is that the values selected for
$\gamma$ are immaterial, so long as the model is flexible enough to assign
sufficient probability mass to each $(\gamma_{\ell}, \gamma_{\ell+1}]$
interval. Since assigning a PPMx prior to $\rho_m$ results in modeling the
latent ordinal scores with a mixture (which is essentially the same data
model arising from marginalizing the random measure in the approach by
\citealt{bao&hanson:2015}), we conclude that the model we specify affords
sufficient flexibility and, therefore, the $\gamma$ can be selected
arbitrarily. In light of this we set $\gamma_1 = 0$, $\gamma_2 = 1/3$,
$\gamma_3 = 2/3$, $\gamma_4=3/3$. The remainder of the model is expressed
hierarchically after introducing latent cluster labels $s_1, \ldots, s_n$
\begin{align} \label{latentModel}
Z_i \mid  \bm{\mu}^{\star}, \bm{\sigma}^{2\star}, s_i & \sim N(\mu^{\star}_{s_i}, \sigma^{2\star}_{s_i}), \nonumber\\
(\mu^{\star}_j, \sigma^{\star}_j) & \sim N(\mu_0, \sigma^2_0) \times Unif(0,A), \\
Pr(\rho \mid  \bm{X}) & \propto \prod_{j=1}^k c(S_j) g(\bm{X}^{\star}_j),\nonumber
\end{align}
where $c(S_j)$ and $g(\bm{X}^{\star}_j)$ are set to functions detailed in
Section \ref{drpm}. Further, we assume $\mu_0 \sim N(0, 10^2)$, $\sigma_0
\sim Unif(0,10)$, and set $A=0.1$.  

This model was fit to the
osteonecrosis data by collecting 1000 MCMC iterates after discarding the
first 10,000 as burn-in and thinning by 15. Afterwards, we identifyed association rules for each MCMC iterate of  $\rho_m$. Covariate values for those variables
not identified in the association rule are set to the overall median.
Testing \eqref{BNP2} was done by using \cite{chen&hanson}'s method based
on 500 permutations and 100 posterior predictive draws.  This testing
procedure was replicated five times with a unique set of 100 random
posterior predictive draws and the average $p$-value is reported. (We
stress again that $p$-values are used only as a validation summary and no
notion of Type I error is implied.) As in the simulation study, each
continuous covariate was dichotomized such that an equal number of
observations belonged to each interval when employing association rules.
Also, we explored the impact that trichotomizing continuous covariates
might have on the  RAID exploratory procedure. Results are provided in the
online supplementary material.

\begin{figure}[htbp]
\hspace{-1 cm}
\includegraphics[scale=0.45]{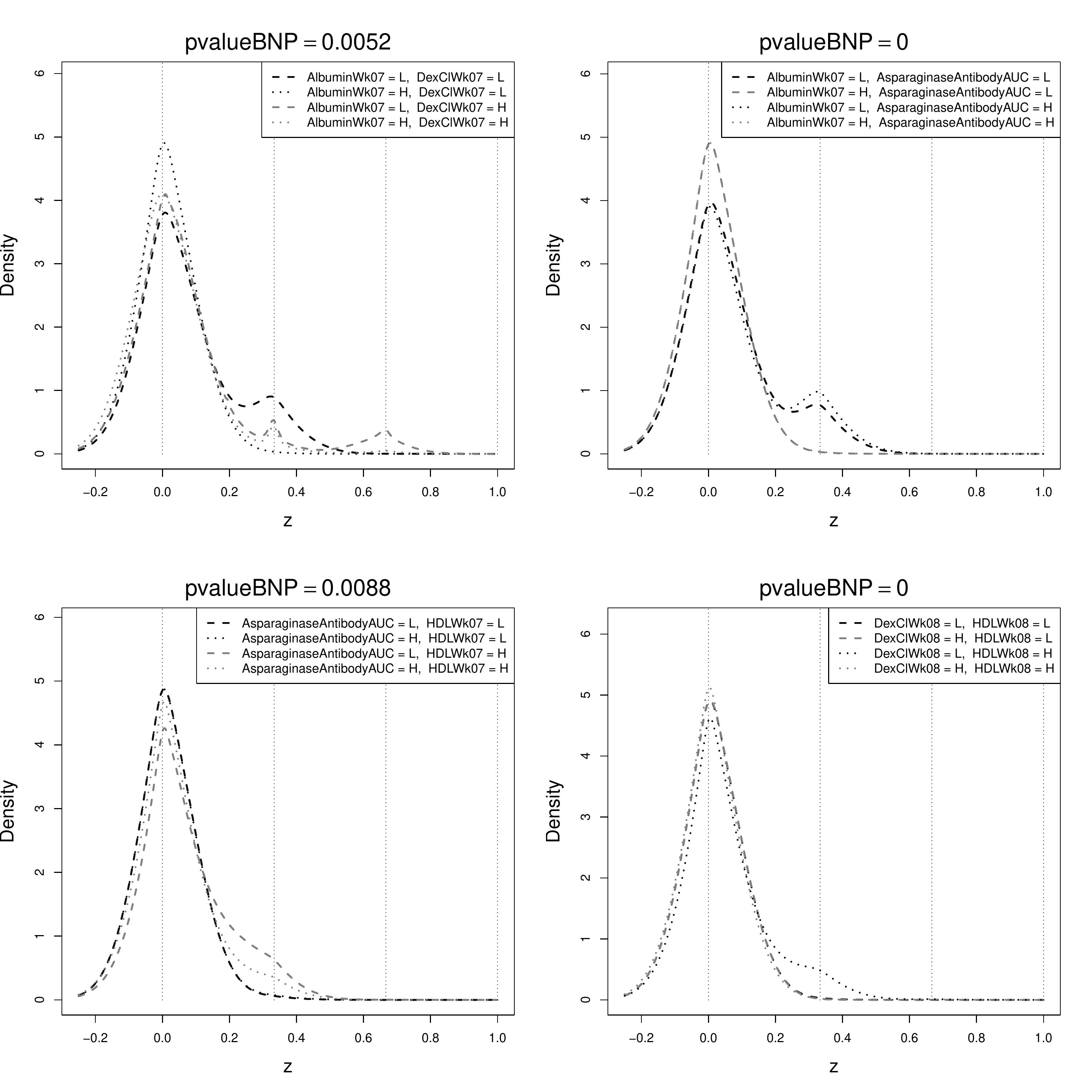}
\caption{Posterior predictive densities for the latent variable $z$ and used in
hypothesis test \eqref{BNP2} for the first four association rules in Table
\ref{aruls2} that do not include {LowRisk}.  The vertical lines
correspond to $\gamma_1, \ldots, \gamma_4$} \label{UniPostPredPC}
\end{figure}

Since the study was conducted specifically to explore if dexamethasone
clearance (DexCl) and/or asparaginase (AsparaginaseAntibodyAUC) interact
with other physiological measurements, we restrict attention to those
association rules that contain at least one of these two covariates.
Results for interactions that appeared in at least 50\% of the MCMC
iterates are provided in Table \ref{aruls2}.  The column ``{\it Pr}''
corresponds to the percent of MCMC iterates that identified the particular
association rule, ``Supp.'' is the support, ``Conf.'' the confidence,
$|S|$ corresponds to average cluster size from which association rule was
identified and  ``$p$-value'' indicates the result of testing
\eqref{BNP2}. Details associated with the ``Prior\#'' column are provided
in Section \ref{priorSensitivity}. Notice that {DexClWk07} interacts
with {LowRisk} and {AlbuminWk07} while {DexClWk08} interacts
with {HDLWk08}. On the other hand, {Asparaginase} seems to
interact with {AlbuminWk07}, {HDLWk07}, and {LDLWk08}. Figure
\ref{UniPostPredPC} contains the posterior predictive densities
corresponding to the first four association rules listed in Table \ref{aruls2}
that do not contain {LowRisk}. Interestingly, note that
the interaction between {AlbuminWk07} and {
AsparaginaseAntibodyAUC} influences the shape of the predictive densities
verifying that there is indeed an interaction between them and the
interaction seems to impact the shape of the response distribution (as
evidenced by changes in shape of the predictive distributions).

\begin{table}[ht]
\caption{Association rules from the osteonecrosis data with  continuous covariates being dichotomized. Here
only association rules that contain either {DexClwk07}, {
DexClwk08}, or {AsparaginaseAntibodyAUC} are considered}
\resizebox{\textwidth}{!}{
\begin{tabular}{ r@{$\ \Leftrightarrow \ $} l cccccr}
\toprule
  \multicolumn{2}{c}{Association Rule} & $Pr$ & Supp. & Conf.  &$|S|$ & $p$-value & Prior\#\\
\midrule
  DexClWk07 & LowRisk 						& 1.00 & 0.62 & 0.91 & 43.44 & 0.00 & 16 \\
  DexClWk07 & AlbuminWk07 					& 0.99 & 0.58 & 0.85 & 39.65 & 0.01 & 15 \\
  AsparaginaseAntibodyAUC & AlbuminWk07 		& 0.93 & 0.84 & 0.95 & 39.35 & 0.00 & 14 \\
  AsparaginaseAntibodyAUC & HDLWk07 			& 0.81 & 0.72 & 1.00 & 10.40 & 0.01 & 7 \\
  AsparaginaseAntibodyAUC & LowRisk 			& 0.81 & 0.87 & 0.98 & 40.55 & 0.18 & 17 \\
  AsparaginaseAntibodyAUC & LDLWk08 			& 0.55 & 0.53 & 0.88 & 15.71 & 0.00 & 6 \\
  DexClWk08 & HDLWk08 					& 0.53 & 0.51 & 0.88 & 16.36 & 0.00 & 8 \\
  DexClWk08 & AlbuminWk08 					& 0.50 & 0.83 & 0.98 & 10.66 & 0.85 & 3 \\
   \bottomrule
\end{tabular}
}
\label{aruls2}
\end{table}

Being able to interpret results at the latent level is not always
straightforward. Thus it would be appealing to determine how interactions
influence the risk of osteonecrosis. Table \ref{ordProb1} contains
predictive probabilities associated with each osteonecrosis grade for
specific levels of covariates found in interactions displayed Figure
\ref{UniPostPredPC} that contain {DexCl}. It appears that low level of
{DexCl} results in higher risk of grade 2 or higher osteonecrosis, but
the exact risk depends on the levels of the other covariates ({
AlbuminWk07} and {HDLWk08}). The combination of low {DexClWk07}
and high {AlbuminWk07} results in the highest risk of osteonecrosis.
This leads one to hypothesize that a three-way interaction between {
Albumin}, {HDL}, and {DexCl} may be present. We explored this by
testing a version of \eqref{BNP2} that includes posterior predictive
densities for all possible combinations of the three covariates (eight in
total). Doing this resulted in rejecting the null (with a $p$-value of 0)
that all densities are equal and hence we conclude that a three-way
interaction exists. More details are provided in the online supplementary
material. 

The online supplementary material also provides details  on how
the RAID procedure can accommodate the temporal structure that exists
among the covariates through the PPMx prior. This is done by considering a
multivariate similarity. As is seen in the supplementary material,
including temporal dependence in the partition prior affects partition
configuration and as a result interactions that are identified by the
RAID procedure.

\begin{table}[htp]
\caption{Posterior probabilities for each severity grade of
osteonecrosis corresponding to two association rules found in Table
\ref{aruls2}}
\begin{center}
\begin{tabular}{l |ccccc}
\multicolumn{1}{c}{} & \multicolumn{5}{c}{Osteonecrosis Grade} \\ \cmidrule(r){2-6}
\multicolumn{1}{c}{Covariates} & 0 & 1 & 2 & 3 & 4  \\ \midrule
AlbuminWk07 - L, DexClWk07 - L 	 & 0.124 & 0.818 & 0.058 & 0.000 & 0.000 \\
AlbuminWk07 - L, DexClWk07 - H 	 & 0.193 & 0.805 & 0.002 & 0.000 & 0.000 \\
AlbuminWk07 - H, DexClWk07 - L 	 & 0.282 & 0.643 & 0.061 & 0.014 & 0.000 \\
AlbuminWk07 - H, DexClWk07 - H 	 & 0.414 & 0.569 & 0.017 & 0.000 & 0.000 \\    \midrule
DexClWk08 - L, HDLWk08 - L 		 & 0.200 & 0.799 & 0.001 & 0.000 & 0.000 \\
DexClWk08 - L, HDLWk08 - H 		 & 0.195 & 0.805 & 0.000 & 0.000 & 0.000 \\
DexClWk08 - H, HDLWk08 - L 		 & 0.287 & 0.682 & 0.030 & 0.001 & 0.000 \\
DexClWk08 - H, HDLWk08 - H		 & 0.330 & 0.667 & 0.003 & 0.000 & 0.000 \\
\bottomrule
\end{tabular}
\end{center}
\label{ordProb1}
\end{table}%

\subsection{Sensitivity to Prior Specification}\label{priorSensitivity}

Even though we developed the RAID procedure to be as automatized as
possible (after standardizing covariates and possibly the response),  ``tuning'' parameters that may affect the list of possible
interactions exist. Those that are most influential are prior parameter
specifications that have direct impact on $\pi(\rho_m\mid \bm{Y}, \bm{X})$. In
Section \ref{sim.studies} we mentioned that for  model \ref{modelNormal}
these are $A$ and $k_0$, as they regulate the influence that $\bm{Y}$ and
$\bm{X}$ have on cluster formation.  Note that as $A \rightarrow 0$,
clusters become more heterogeneous with respect to $\bm{Y}$ while as $k_0
\rightarrow 0$ clusters become more heterogeneous with respect to
$\bm{X}$.  To explore sensitivity to these three prior specifications, we
ran the RAID procedure for the osteonecrosis data based on $A \in \{0.1,
1, 10\}$ and $k_0 \in \{0.1, 1, 10\}$.  Since the cohesion function also
has an effect on $\pi(\rho_m\mid \bm{Y}, \bm{X})$, we ran the RAID procedure
using $c(S_j) = 1$ (a uniform type cohesion function) in addition to
$c(S_j) = M\times(|S|-1)!$. This resulted in 18 prior configurations and as a result 18 runs of the RAID procedure. 
On average it took 15 minutes to employ the RAID to the 18 prior configurations. We provide tables that list
the potential interactions for each of the 18 runs in the
online supplementary material. Here we briefly summarize the results.

Values in the column labeled ``Prior\#''  in Table  \ref{aruls2} are the number of prior configurations that detected the 
corresponding interaction.  Counting up the total number of unique interactions detected across the 18 runs of the RAID procedure resulted in 23 unique
interactions.  Each of the interactions listed in Table \ref{aruls2} were among the unique 23.   There was a general agreement in the interactions detected among the 18 runs with four of the interactions in Table \ref{aruls2}  appearing in the majority of the prior configurations.  Overall, it seems that results are fairly
robust to prior specifications.

Lastly, as a type of follow-up analysis, using the {\tt polr} function
found in the {\tt MASS} {\tt R}-package (\citealt{MASS}), we fit an ordered
logistic regression model that included all  eight two-way interactions
listed in Table \ref{aruls2} and their corresponding main effects along
with the three-way interaction detailed in the online supplementary
material. (We remark that fitting a saturated ordered
logistic regression model with all two-way interactions for these data is
not possible.) This model did provide some evidence to suggest that an
interaction between {DexClWk08} and {HDLWk08} along with {\tt
AsparaginaseAntibodyAUC} and {HDLWk07} was present in addition to the
three-way interaction, but failed to identify any of the other
interactions. This suggests that the added flexibility of
our exploratory approach may be useful to detect interactions that
would not be possible under more restricted alternatives.

\section{Conclusions}\label{sect:disc}

We have detailed an exploratory procedure that in a general way is able to
identify interactions. The definition of interaction that we espouse is
more general than that typically associated with linear models. Here we
conceptualize an interaction as influencing any aspect of the response
distribution rather than focusing solely on the first moment. Further, the
procedure is able to identify interactions without {\it a priori}
information regarding which (multiplicative) effects to include in a data
model. This was seen when applying the procedure to the osteonecrosis data
set. In the application, relationships that were known to exist (e.g., Low
{DexCl} and higher risk of osteonecrosis) were identified without
considering them explicitly in the procedure. In addition, a three-way
interaction, also not explicitly included in the data model, was
highlighted (see supplementary material). This interaction was not
previously known by the investigators, and its scientific relevance has
yet to be determined.

We also demonstrated that the procedure can be easily employed regardless
of the type of response being measured. Indeed, the procedure is
essentially ``data model free'' as it can be employed regardless of the
data model so long as some type of random partition model is employed.
Note, however,  that our exploratory procedure only identifies potential
interactions. That is, it does not estimate their effect. If this is
desired, then any number of statistical procedures that include the
specific covariates identified by our procedure can be employed (i.e.,
some form of regression).

One limitation of the RAID procedure (which is more a MCMC limitation) is
that computation associated with the PPMx does not tend to scale well as
$p$ and/or $n$ grows. Future research will be dedicated to implementing
methods in the RAID procedure that permits it to scale well.

Finally, we provide some general recommendations when employing the RAID
procedure. For the first step we recommend using the PPMx as the RPM. To
automate the prior specification of model \ref{modelNormal}, we suggest
standardizing the response and covariates and setting $A=0.5$ and $k_0 =
0.5$.  This   essentially puts equal weight on $\bm{Y}$ and $\bm{X}$ when
forming clusters. Decisions associated with step two determine how big a
``net'' one wants to cast when searching for interactions with the
understanding that a larger ``net'' could result in more false positives.
Decreasing support and confidence will cast a larger ``net'' as will
decreasing the number of MCMC iterates in which an association must
appear. Informal explorations led us to set support to 0.25,  confidence
of 0.5, and that a particular association rule should appear in 50\% of
MCMC iterates before being formally considered.    For step three we
suggest using 100 posterior predictive draws when comparing predictive
densities and a cut-off of 0.01 for ``$p$-values'' if the
\cite{chen&hanson} method is used. This is somewhat arbitrary but worked
to identify interactions well in the simulations. Lastly, we recommend
using posterior predictive densities to visualize the interactions. Doing
so will highlight how the interaction is affecting the response
distribution, be it a shift in expectation as in Figure \ref{meanChange} or
a change in shape as in Figure \ref{UniPostPredPC}. If it is determined
that the interaction affects the expectation of the response distribution,
then a follow-up analysis to quantify interaction effects can be carried
out using generalized linear models. If the interaction affects the
response distribution in some other way, then formal additional analysis
is not available and quantifying the interaction effects will requiring
other (perhaps ad-hoc) approaches similar to our approach with the
osteonecrosis data set.

\section*{Acknowledgements}

We thank all the reviewers for comments and
suggestions that helped us to greatly improve this manuscript.  We thank Drs. Yuhui Chen and Timothy Hanson for kindly
sharing code that carries out the P\'olya Tree permutation test. Garritt
L. Page was partially supported by the Basque Government through the BERC 2018-2021
program, by the Spanish Ministry of Science, Innovation and Universities
through BCAM Severo Ochoa accreditation SEV-2017-0718. Fernando A.
Quintana was supported by grant FONDECYT 1180034 and by Millennium Science
Initiative of the Ministry of Economy, Development, and Tourism, grant
``Millenium Nucleus Center for the Discovery of Structures in Complex
Data''. Gary L. Rosner was partially funded by grants GM092666 and
P30CA006973.

\bibliographystyle{imsart-nameyear}
\bibliography{../reference}

\end{document}


\onehalfspace

\title{Supplementary Material: Discovering Interactions Using Covariate Informed Random Partition Models}
\author{ Garritt L. Page \\ Department of Statistics \\ Brigham Young University \\  Basque Center for Applied Mathematics \\ page@stat.byu.edu
         \and	Fernando A. Quintana \\ Departamento de Estad\'{i}stica \\ Pontificia Universidad Cat\'{o}lica de Chile \\ quintana@mat.uc.cl
         \and	Gary L. Rosner \\ Department of Biostatistics \\ John Hopkins University \\ grosner1@jhmi.edu
}

\maketitle

\noindent This document contains supplementary material to the paper
``Discovering Interactions Using Covariate Informed Random Partition
Models''. In Section \ref{simstudy} we provide more details associated
with the simulation studies conducted in Sections
\ref{sim.study.toyexample} and \ref{sim.study.2} of the main article.
Section \ref{osteonecrosis} provides more results from the osteonecrosis
study in Section \ref{InteractionsInStudy} of the main article.  These
include results when continuous covariates are trichotomized, a three-way
interaction between {\tt AlbuminWk07}, {\tt DexClWk07}, and {\tt HDLWk08}
is explored, a multivariate similarity to accommodate temporal dependence
in covariates is employed,  and all results from the sensitivity study.

\section{Additional Details Associated with Simulation Studies}\label{simstudy}

We first provide more insight from the simulation study in Section \ref{sim.study.toyexample} and
then from that of Section \ref{sim.study.2}.

\subsection{Details of Competitors Included in Both Simulation Studies}

As mentioned in the main article, in a simulation study setting the
decision of whether an interaction is present or not
needs to be made automated. All procedures need some oversight as it is
not always clear if the automated decision is correct or not. To this end
we carefully detail the process used for each procedure.
\begin{itemize}
\item {\bf Linear Model}: Using the {\tt lm} function in {\tt R}, we
    fit a saturated model with all main effects and two-way
    interactions. An interaction was deemed
    ``detected'' if the frequentist $p$-value was less than 0.01.

\item {\bf Spike \& Slab}: Using the {\tt spikeslab} package, similar
    to the linear model we fit a saturated model with all main effects
    and two-way interactions.  All interaction effects that have a
    non-zero ``gnet'' value are considered to be ``real''
    interactions.  This follows the suggestion outlined in
    \cite{ish&kogalur&rao2010}.   The ``gnet'' value is defined as the
    model in a particular path solution that minimizes the AIC
    criterion.   See \cite{ish&kogalur&rao2010} or
    \cite{ishwaran&rao:2005} for more details.

\item {\bf hgLASSO}: Using the {\tt glinternet} package we fit a
    grouped LASSO procedure with the shrinkage parameter being
    estimated using 10-fold cross validation.  More specifically, the
    {\tt glinternet.cv} function was used, with the ``lambda''
    argument set to 10.  Then the ``lambda'' parameter that
    corresponded to  cross-validation error estimate below the minimum
    plus one standard deviation was used.  On occasion, this procedure
    produced an error (particularly in the case when no interactions
    were explicitly included in the data generating mechanism).  When
    this occurred, we simply skipped the data iteration.

\item {\bf BART}: We used the {\tt bartMachine} package to fit the
    BART model and then the function {\tt interaction\_investigator}
    identify potential interactions.  Covariates are said to interact
    in a given tree ``only if they appear together in a contiguous
    downward path from the root node to a terminal node''.  Summing
    across trees and MCMC iterations produces a total number of
    interactions for each pair of variables.  This value is used to
    rank the interactions on their ``importance''.   See
    \cite{kapelner&bleich:2016} for specific details.  No direction is
    given in  \cite{kapelner&bleich:2016} on a cut-off value that
    could be used to classify an interaction as ``real'' or not.  Thus, we devised
    and used the following procedure.  Interactions are ordered by their
    ``total'' and a cut-off is determined by a 50\% decrease in the
    ``total'' value.  All interactions above this threshold are
    considered ``real''.

\item{\bf XGBoost}: We followed the vignette found in \cite{EXI:int}
    to highlight potential interactions using XGBoost. However, just
    like the BART procedure, little guidance is given on when to
    classify potential interactions as noise. To this end, we employed
    the same strategy as in the case of the BART
    procedure. That is, we ordered interactions according to their
    ``sumGain'' and a cut-off is determined by a 50\% decrease in the
    ``sumGain'' value. All interactions above this threshold are
    considered ``real''. It was not uncommon for the {\tt xgboost}
    function to produce an error in data generating scenarios that
    included interactions that did not depend on a shift of
    expectation. When this occurred we simply skipped the data
    iteration.

\end{itemize}

\subsection{Additional Details for the Simulation Study Based on a Toy Example}

Since only a small number of two-way association rules in the simulation
study exist, we are able to provide results for each one which facilitates
illustrating each step of the exploratory approach. The association rules
can be found in Table \ref{aRulesummary} under the header ``arules''. The
column ``$Pr$'' contains the percent of MCMC iterates (out of 1000) that
identified the particular association rule averaged across the 1000
synthetic data sets. Notice that under $f_0(Y|\bm{X})$  no association
rule is favored over any other with regards to its chances of being
selected. This is to be expected as $\bm{X}$ in no way influences $Y$.
Contrast that to when $\bm{X}$ does influence $Y$, (i.e., under
$f_1(Y|\bm{X})$, $f_2(Y|\bm{X})$, and $f_3(Y|\bm{X})$). In these cases,
association rules that include both $X_1$ and $X_2$ are much more highly
favored in terms of the chances of being detected compared to those that
in any way contain $X_3$. More concretely, the association rules that
include $X_1$ and $X_3$ all have very low chances of being detected while
those that include $X_2$ and $X_3$ have a higher chance but still much
lower than those that include both $X_1$ and $X_2$ (approximately 30\%
compared to 80\%).
\begin{center}
\begin{table}[ht]
\caption{Summary of characteristics for each association rule
corresponding to simulation study results.  See the text for an explanation. }
\resizebox{17cm}{!}{\begin{tabular}{l ccc | ccc | ccc | ccc }
\toprule
 & \multicolumn{3}{c}{$f_0(Y|\bm{X})$} & \multicolumn{3}{c}{$f_1(Y|\bm{X})$} & \multicolumn{3}{c}{$f_2(Y|\bm{X})$} &
 \multicolumn{3}{c}{$f_3(Y|\bm{X})$} \\  \cmidrule(lr){2-4} \cmidrule(lr){5-7}  \cmidrule(lr){8-10} \cmidrule(lr){11-13}
  \multicolumn{1}{c}{arules} & $Pr$ & Size & \#dat & $Pr$ & Size & \#dat & $Pr$ & Size & \#dat & $Pr$ & Size & \#dat \\
\midrule
 $\{X_1=0\}$  $\Rightarrow$ $\{X_2=0\}$ 	& 0.11 & 36.71 & 100 & 0.07 & 56.31 & 100 & 0.90 & 30.51 & 100 & 0.87 & 60.86 & 100 \\
 $\{X_1=0\}$  $\Rightarrow$ $\{X_2=1\}$ 	& 0.10 & 35.56 & 100 & 0.32 & 75.52 & 100 & 0.45 & 88.77 & 100 & 0.42 & 90.15 & 100 \\
 $\{X_1=0\}$  $\Rightarrow$ $\{X_3=0\}$ 	& 0.08 & 48.65 & 100 & 0.02 & 19.82 & 99.4 & 0.03 & 17.02 & 99.9 & 0.02 & 20.57 & 99.2 \\
 $\{X_1=0\}$  $\Rightarrow$ $\{X_3=1\}$ 	& 0.07 & 44.70 & 100 & 0.02 & 19.46 & 99.2 & 0.03 & 17.13 & 99.9 & 0.02 & 20.79 & 99.8 \\
 $\{X_1=1\}$  $\Rightarrow$ $\{X_2=0\}$ 	& 0.11 & 36.96 & 100 & 0.07 & 55.79 & 100 & 0.53 & 32.23 & 100 & 0.37 & 35.22 & 100 \\
 $\{X_1=1\}$  $\Rightarrow$ $\{X_2=1\}$ 	& 0.11 & 35.52 & 100 & 0.33 & 77.38 & 100 & 0.20 & 78.94 & 100 & 0.26 & 69.67 & 100 \\
 $\{X_1=1\}$  $\Rightarrow$ $\{X_3=0\}$ 	& 0.06 & 46.41 & 100 & 0.02 & 19.85 & 99.1 & 0.05 & 19.58 & 100 & 0.08 & 27.48 & 100 \\
 $\{X_1=1\}$  $\Rightarrow$ $\{X_3=1\}$ 	& 0.07 & 45.47 & 100 & 0.02 & 19.96 & 98.6 & 0.05 & 19.67 & 100 & 0.09 & 27.53 & 100 \\ 			\midrule
 $\{X_2=0\}$  $\Rightarrow$ $\{X_1=0\}$ 	& 0.08 & 42.91 & 100 & 0.97 & 120.58 & 100 & 0.38 & 18.08 & 100 & 0.80 & 59.99 & 100 \\
 $\{X_2=0\}$  $\Rightarrow$ $\{X_1=1\}$ 	& 0.08 & 42.90 & 100 & 0.97 & 120.67 & 100 & 0.23 & 25.48 & 100 & 0.84 & 61.36 & 100 \\
 $\{X_2=0\}$  $\Rightarrow$ $\{X_3=0\}$ 	& 0.08 & 43.93 & 100 & 0.01 & 18.14 & 67.7 & 0.11 & 20.48 & 100 & 0.05 & 25.15 & 100 \\
 $\{X_2=0\}$  $\Rightarrow$ $\{X_3=1\}$ 	& 0.08 & 43.71 & 100 & 0.01 & 17.39 & 66.5 & 0.11 & 20.61 & 100 & 0.05 & 25.20 & 100 \\
 $\{X_2=1\}$  $\Rightarrow$ $\{X_1=0\}$ 	& 0.08 & 42.02 & 100 & 0.06 & 26.65 & 100 & 0.08 & 26.05 & 100 & 0.10 & 45.48 & 100 \\
 $\{X_2=1\}$  $\Rightarrow$ $\{X_1=1\}$ 	& 0.08 & 42.89 & 100 & 0.06 & 26.32 & 100 & 0.05 & 39.55 & 100 & 0.10 & 44.61 & 100 \\
 $\{X_2=1\}$  $\Rightarrow$ $\{X_3=0\}$ 	& 0.08 & 44.29 & 100 & 0.05 & 26.12 & 100 & 0.05 & 26.36 & 100 & 0.06 & 28.35 & 100 \\
 $\{X_2=1\}$  $\Rightarrow$ $\{X_3=1\}$ 	& 0.07 & 41.69 & 100 & 0.06 & 26.15 & 100 & 0.05 & 25.75 & 100 & 0.06 & 28.36 & 100 \\ 			\midrule
 $\{X_3=0\}$  $\Rightarrow$ $\{X_1=0\}$ 	& 0.07 & 53.10 & 99.8 & 0.01 & 16.57 & 95.5 & 0.02 & 16.71 & 99.4 & 0.01 & 18.46 & 95 \\
 $\{X_3=0\}$  $\Rightarrow$ $\{X_1=1\}$ 	& 0.06 & 51.65 & 99.7 & 0.01 & 17.21 & 95.5 & 0.04 & 26.42 & 99.9 & 0.09 & 39.57 & 100 \\
 $\{X_3=0\}$  $\Rightarrow$ $\{X_2=0\}$ 	& 0.05 & 63.40 & 98.2 & 0.00 & 10.98 & 31.6 & 0.35 & 31.30 & 100 & 0.06 & 38.14 & 99 \\
 $\{X_3=0\}$  $\Rightarrow$ $\{X_2=1\}$ 	& 0.05 & 62.93 & 99 & 0.27 & 102.76 & 100 & 0.24 & 104.69 & 99.9 & 0.22 & 96.82 & 100 \\
 $\{X_3=1\}$  $\Rightarrow$ $\{X_1=0\}$ 	& 0.05 & 50.60 & 99.9 & 0.01 & 16.38 & 95.2 & 0.02 & 16.77 & 99.5 & 0.01 & 18.76 & 95 \\
 $\{X_3=1\}$  $\Rightarrow$ $\{X_1=1\}$ 	& 0.07 & 52.88 & 100 & 0.01 & 16.48 & 93.6 & 0.04 & 26.20 & 99.9 & 0.09 & 39.42 & 100 \\
 $\{X_3=1\}$  $\Rightarrow$ $\{X_2=0\}$ 	& 0.05 & 63.96 & 99.1 & 0.00 & 11.53 & 27.9 & 0.34 & 31.06 & 100 & 0.06 & 37.46 & 98.9 \\
 $\{X_3=1\}$  $\Rightarrow$ $\{X_2=1\}$ 	& 0.05 & 59.35 & 99.6 & 0.26 & 100.68 & 100 & 0.24 & 103.72 & 100 & 0.25 & 100.64 & 100 \\    \bottomrule
\end{tabular}}
\label{aRulesummary}
\end{table}
\end{center}
It is interesting  to point out that under  $f_3(Y|\bm{X})$  the procedure
seems to be able to identify the correct association with more certainty
compared to other scenarios (at least for this data generating scenario).
These same trends are further corroborated by the ``Size'' column. Values
in this column provide the cluster size from which each association rule
was identified averaged over the 1000 MCMC iterates and 1000 synthetic
datasets (recall each data set is comprised of 500 observations). Notice
that when $\bm{X}$ in no way influences the distribution of $Y$, then
average cluster size is essentially uniform across all association rules.
By way of contrast, cluster sizes on average are bigger for association
rules that include both $X_1$ and $X_2$ compared to those that do not when
$\bm{X}$ influences the distribution of $Y$.  The last column, ``\# dat''
is the percent of synthetic datasets where the association rule was
identified for at least one MCMC iterate. Overall  the take home message
of Table \ref{aRulesummary} is that the procedure is able to identify the
pair of covariates that influence the distribution of $Y$ regardless of
what feature of $Y$'s distribution is being influenced.

\subsection{Additional Details for Simulation Study Based the Osteonecrosis Data}\label{simstudy2}

\begin{figure}[htbp]
\begin{center}
\includegraphics[width=1.0\textwidth]{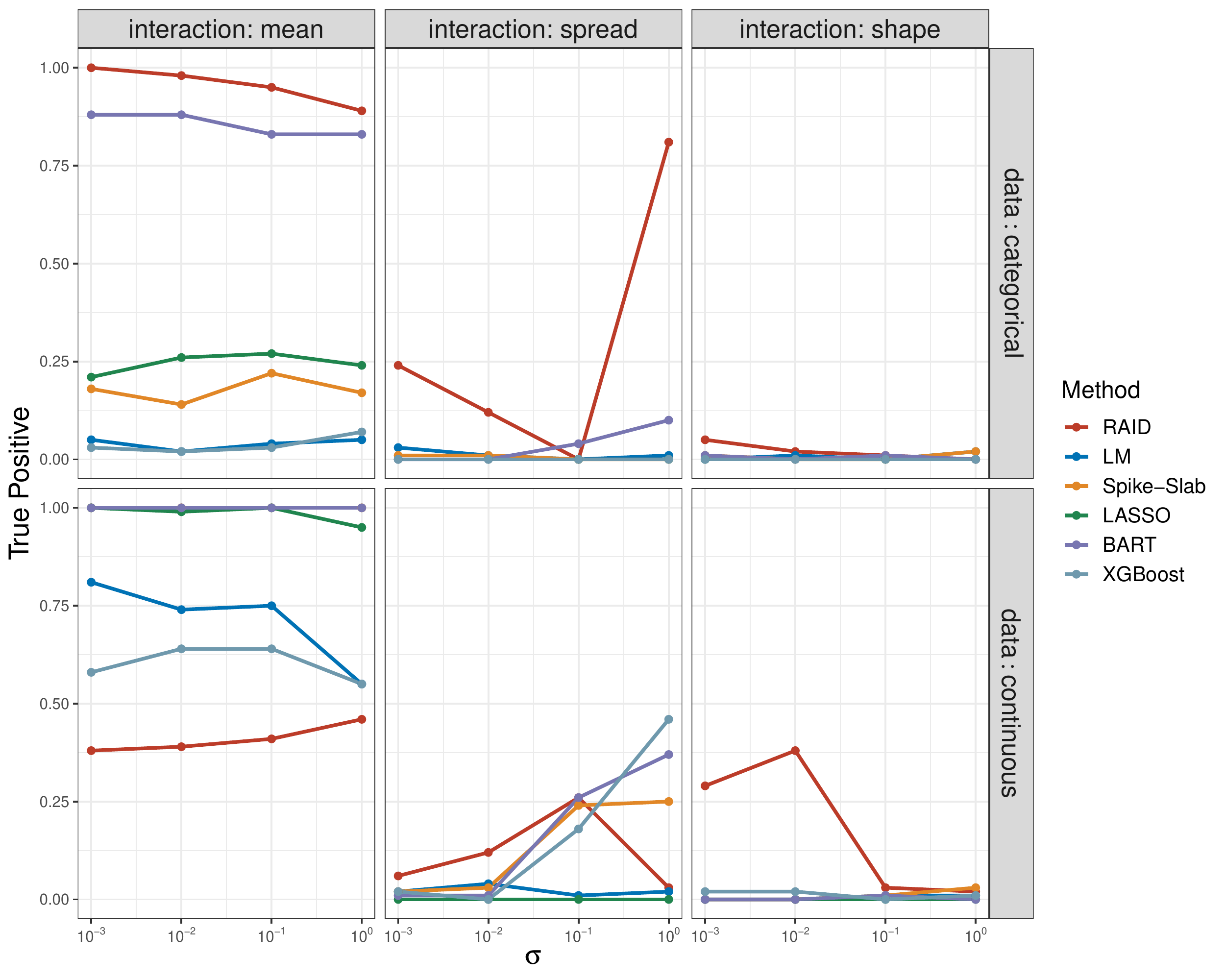}
\caption{Results from the second simulation study when the interaction is exhibited in only
60\% of the population.  The percentage of synthetic datasets for which
the interaction was detected is provided on the $y$-axis. In each figure, the top row corresponds to
an interaction that is based on categorical covariates and the bottom row
to continuous covariates. } \label{simstud2Results3}
\end{center}
\end{figure}

\begin{figure}[htbp]
\begin{center}
\includegraphics[width=1.0\textwidth]{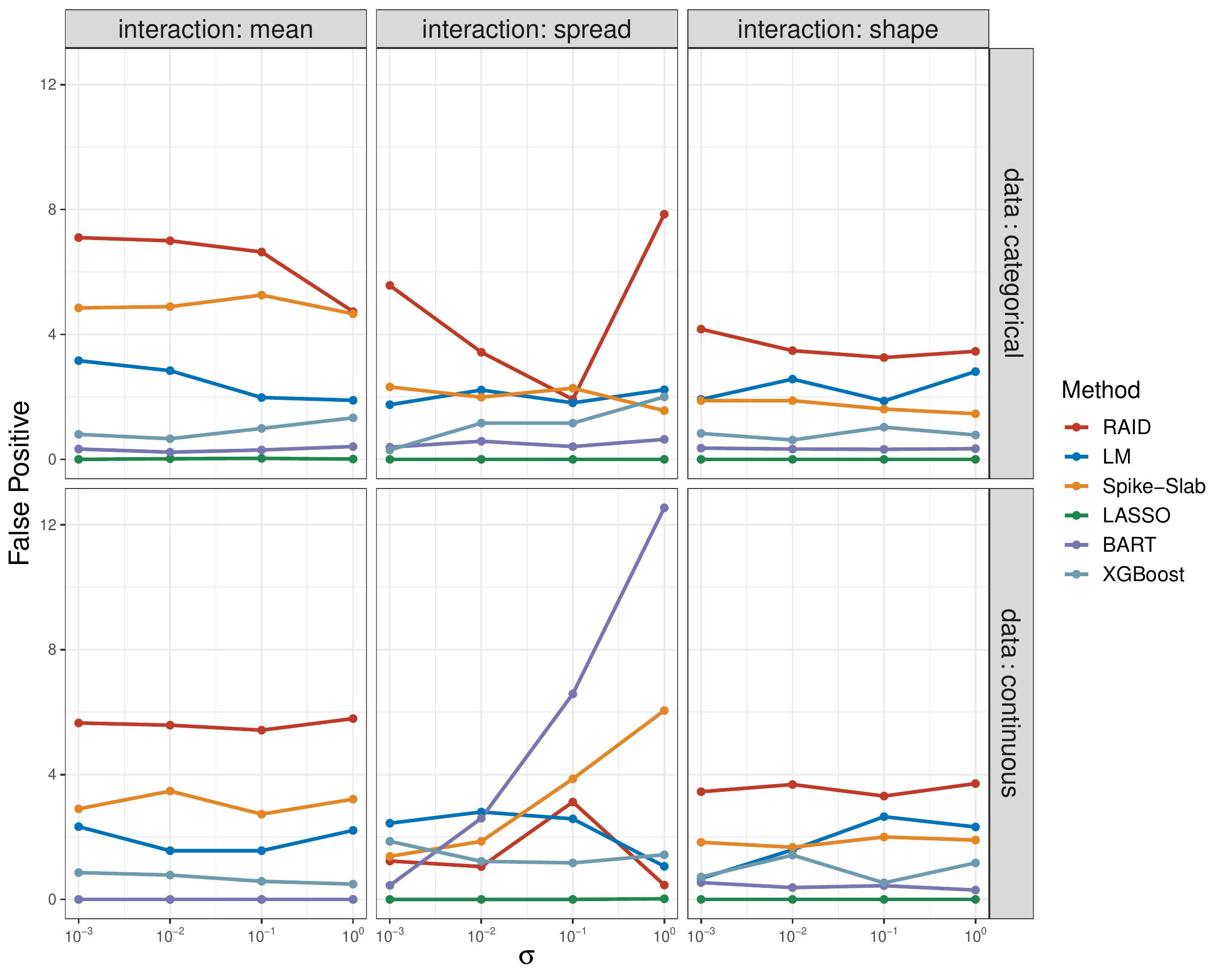}
\caption{Results from the second simulation study when the interaction is exhibited in only
60\% of the population.  The average number of falsely detected
interactions is provided on the $y$-axis.  In each figure, the top row corresponds to an
interaction that is based on categorical covariates and the bottom row to
continuous covariates.   } \label{simstud2Results4}
\end{center}
\end{figure}

\begin{figure}[htbp]
\begin{center}
\includegraphics[width=1.0\textwidth]{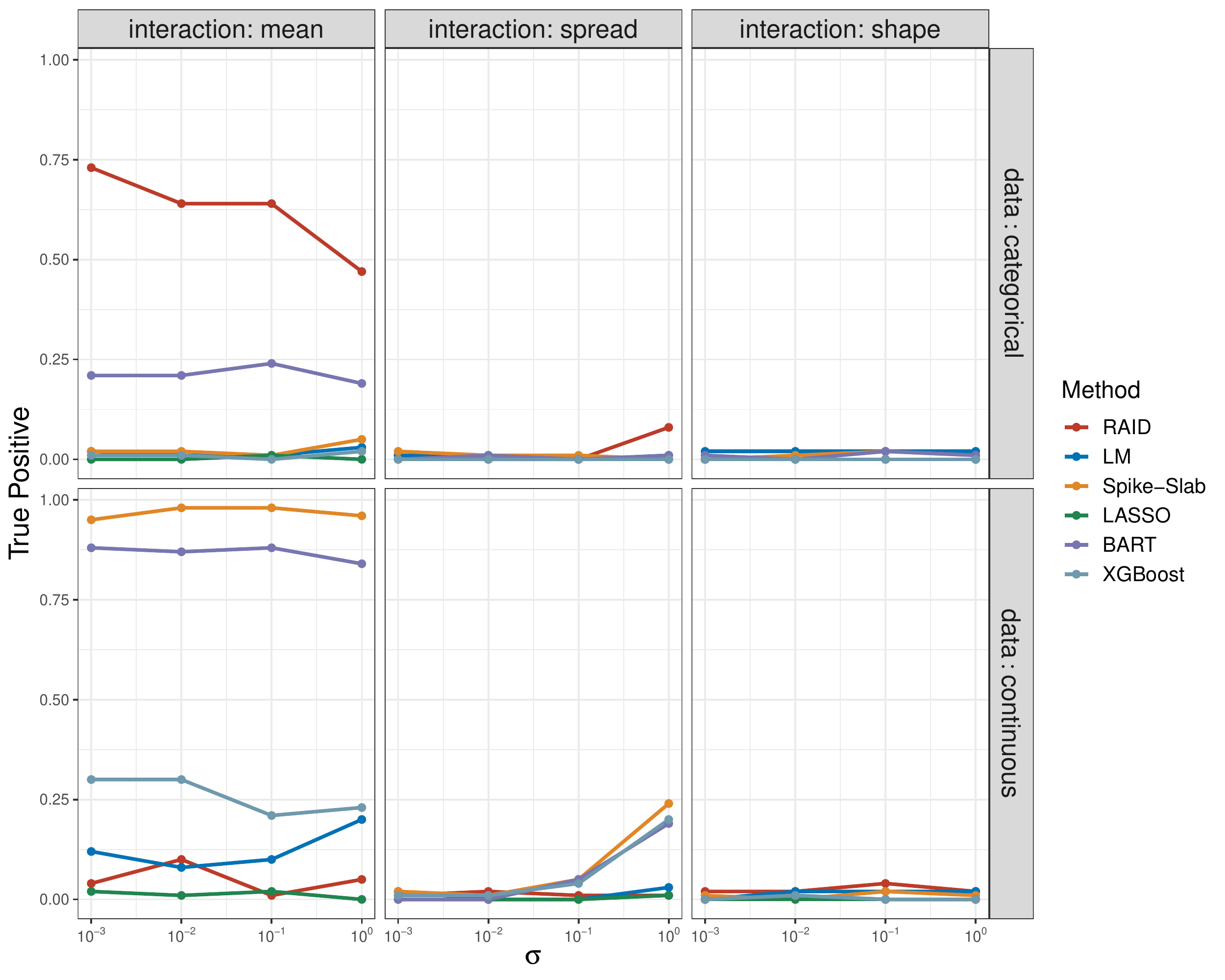}
\caption{Results from the second simulation study when the interaction is exhibited in only
25\% of the population.  The percentage of synthetic datasets for which
the interaction was detected is provided is on the $y$-axis. In each figure, the top row corresponds to
an interaction that is based on categorical covariates and the bottom row
to continuous covariates. } \label{simstud2Results5}
\end{center}
\end{figure}

\begin{figure}[htbp]
\begin{center}
\includegraphics[width=1.0\textwidth]{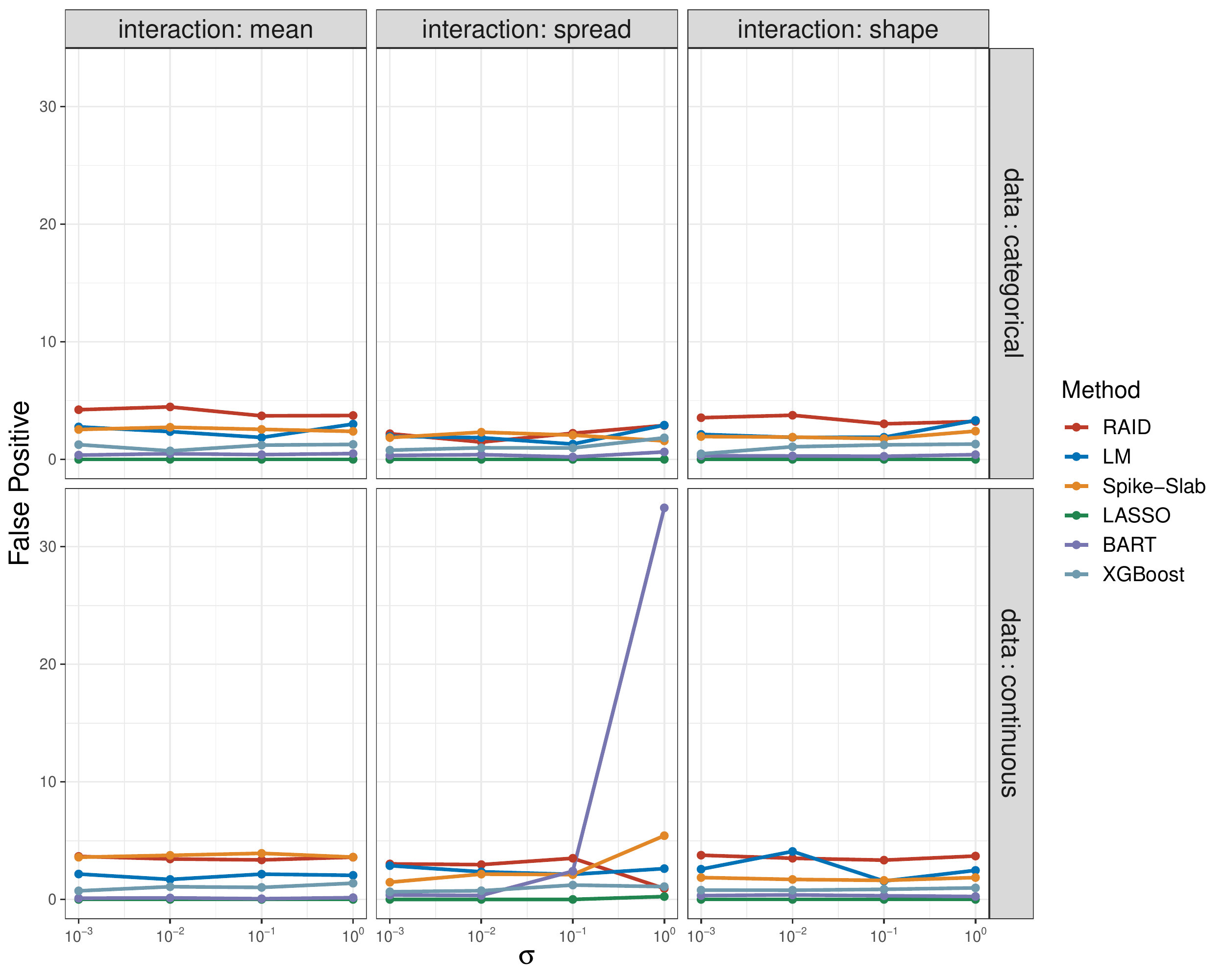}
\caption{Results from the second simulation study when the interaction is exhibited in only
25\% of the population.  The average number of falsely detected
interactions is provided on the $y$-axis.  In each figure, the top row corresponds to an
interaction that is based on categorical covariates and the bottom row to
continuous covariates.   } \label{simstud2Results6}
\end{center}
\end{figure}

In addition to generating data in which the interaction manifests itself
in 100\% of the population, we also considered the cases when only 60\%
and 25\% of the population exhibited an interaction. This was done in the
simulation study in Section \ref{sim.study.2} by randomly selecting 40\%
and 75\% of the generated response values and replacing them with a random
standard normal realization. Results when doing this are provided in
Figures \ref{simstud2Results3} - \ref{simstud2Results6}. As expected all
procedures ability to detect interactions decays as the percent of the
population that exhibit the interaction decreases. With only 60\% of the
population displaying the interaction it seems that the RAID procedure is
the least impacted in that it is able to detect the interaction in all
scenarios except when a two categorical covariates interact in such a way
that the shape of the response distribution changes. The other methods do
well when interaction impacts the mean of the response generally speaking.
When only 25\% of the population exhibit the interaction RAID does well in
detecting interaction when categorical covariates impact mean of response,
while BART and spike \& slab detect it when continuous covariates impact
mean of response. In the other scenarios with only 25\% of population
exhibiting interaction, none of the procedures do well.  In terms of false
positives, the RAID procedure and spike \& slab seem to be more prone to
them. But as mentioned, for RAID this is partly an artifact of the
procedure finding ``main'' effects in addition to ``interactions''.

\section{Further Analysis of the Osteonecrosis Study} \label{osteonecrosis}

\subsection{Using Trichotomized Continuous Covariates in the Osteonecrosis Study}
\label{trich}

Since association rules can only realistically be applied to categorical
variables, it would be natural to question if the discretization  impacts
how association rules highlight covariates that may be interacting.  That
is,  does the  discretization depth influence interaction discovery?
Recall, in the main article, that each continuous covariate was
dichotomized so that both levels (``Low'', ``High'') contained 50\% of the
observations.  It seems reasonable to believe that a finer discretization
of the continuous covariates would discover more ``local'' multiplicative
associations. To explore this, we ran the RAID procedure with
trichotomized continuous covariates so that 1/3 of the observations
belonged to each interval. With continuous covariates being binned into
three groups, the hypothesis in \eqref{BNP2} would become an equality of
nine posterior predictive densities.  As a result, we increased the number
of posterior predictive draws to 300 when carrying out the  P\'olya  tree
test of \cite{chen&hanson}.  The association rules that contain either
{\tt DexCl} or {\tt AsparaginaseAntibodyAUC} are presented in Table
\ref{aruls3}.

\begin{table}[ht]
\center \caption{ ssociation rules from the Osteonecrosis data resulting
from the ordinal data model fit with univariate similarity functions and
continuous covariates being trichotomized (Low, Medium, High)}
\hspace{-0.25 cm}
\begin{small}
\begin{tabular}{ r@{$\ \Leftrightarrow \ $} l ccccc}
\toprule
  \multicolumn{2}{c}{Association Rule} & $Pr$ & Supp. & Conf.  & |S| & $p$-value \\
\midrule
	DexClWk07 & LowRisk   					& 1.00 & 0.43 & 0.91 & 70.14 & 0.00 \\
	DexClWk08 & LowRisk   					& 1.00 & 0.45 & 0.81 & 70.14 & 0.06 \\
	DexClWk07 & TriglyceridesWk08   			& 0.75 & 0.55 & 0.99 & 14.42 & 0.00 \\
 	DexClWk07 & AlbuminWk07 				& 0.66 & 0.64 & 0.85 & 46.40 & 0.00 \\
  	DexClWk08 & Race2  					& 0.61 & 0.39 & 0.71 & 67.55 & 0.35 \\
  	AsparaginaseAntibodyAUC & LowRisk  		& 0.58 & 0.70 & 0.99 & 46.42 & 0.05 \\
 	DexClWk07 & HDLWk07 					& 0.58 & 0.59 & 1.00 & 13.80 & 0.00 \\
  	DexClWk08 & AlbuminWk08 				& 0.51 & 0.57 & 0.98 & 12.90 & 0.49 \\
\bottomrule
\end{tabular}
\end{small}
\label{aruls3}
\end{table}

There are clear similarities between Tables \ref{aruls2} and \ref{aruls3}.
In particular the first association rule in Table \ref{aruls3} is also the
first in Table \ref{aruls2}. That said, there are differences and as a
result, it appears that trichotomizing continuous covariates produces
different potential interactions.  For example, an interaction  between
{\tt TriglyceridesWk08} and {\tt DexClWk07} was detected when continuous
covariates were trichotomized but was not identified in Table
\ref{aruls2}. In fact, {\tt Triglycerides} do not appear in Table
\ref{aruls2}. Overall, it does seem that the manner in which continuous
covariates are discretized does impact which interactions are highlighted.
This of course is not unexpected.

\subsection{Three-Way Interaction}

\begin{figure}[htbp]
\begin{center}
\includegraphics[scale=0.47]{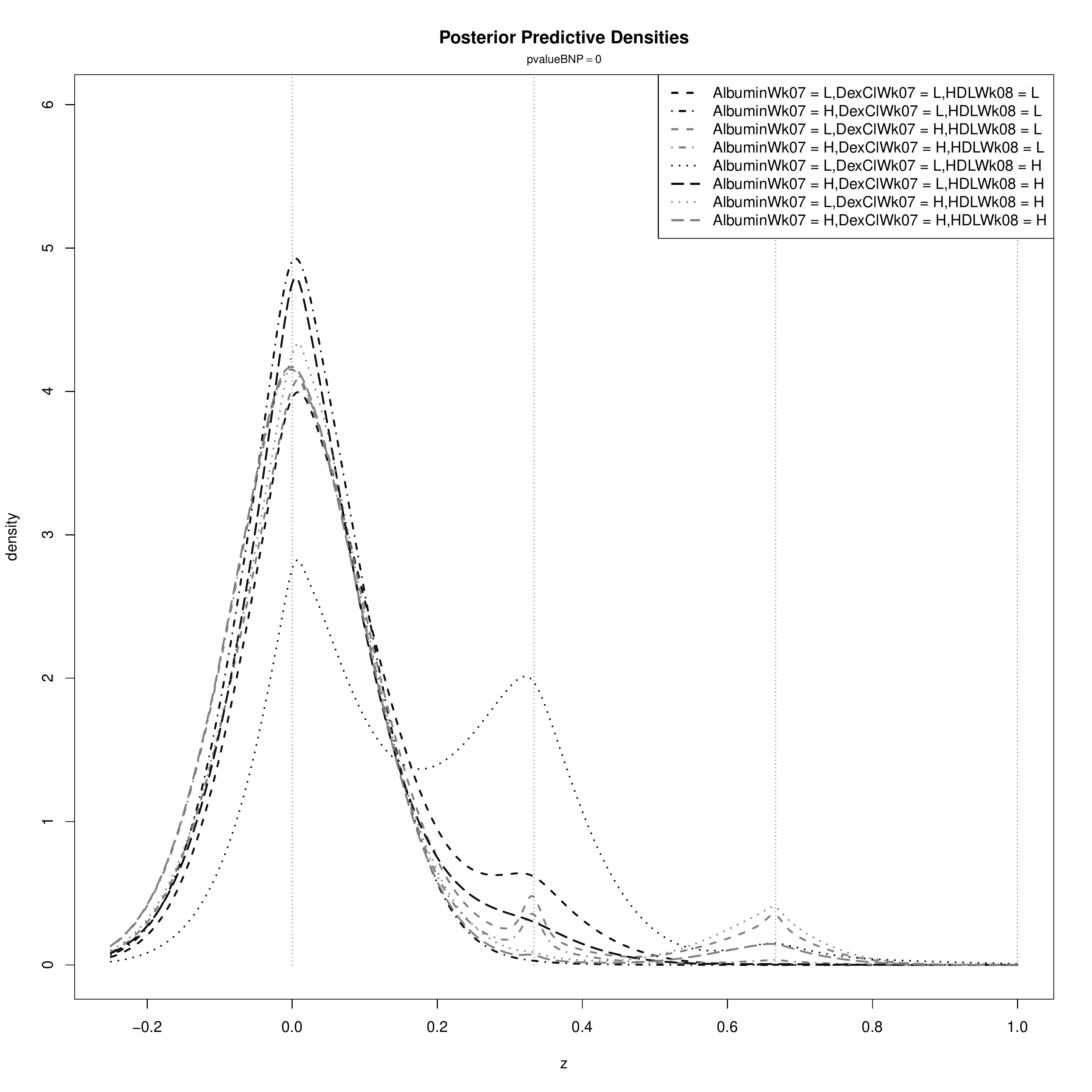}
\caption{Posterior predictive densities for the latent variable $z$ associated with
the eight densities found \eqref{hyp}.  The vertical lines correspond to
$\gamma_1, \ldots, \gamma_4$} \label{threeway}
\end{center}
\end{figure}

The exact hypothesis that was considered when exploring the existence of a
three-way interaction between {\tt DexClWk07}, {\tt HDLWk08}, and {\tt
AlbuminWk07} is (we use $D$, $H$, and $A$ as short hand for {\tt
DexClWk08}, {\tt HDLWk08}, and {\tt AlbuminWk08}, and $\bm{X}$ denotes the
other covariates fixed at their median value)
\begin{align}\label{hyp}
H_0:\, & f(Y_0|\bm{Y}, D=L, H=L, A=L, \bm{X}) = f(Y_0|\bm{Y}, D=L, H=L, A=H, \bm{X}) = \nonumber \\
         & f(Y_0|\bm{Y}, D=L, H=H, A=L, \bm{X}) =  f(Y_0|\bm{Y}, D=L, H=H, A=H, \bm{X}) =\\
	& f(Y_0|\bm{Y}, D=H, H=L, A=L, \bm{X}) = f(Y_0|\bm{Y}, D=H, H=L, A=H, \bm{X}) = \nonumber\\
         & f(Y_0|\bm{Y}, D=H, H=H, A=L, \bm{X}) =  f(Y_0|\bm{Y}, D=H, H=H, A=H, \bm{X}).  \nonumber
\end{align}
We used the same testing procedure as before to test this hypothesis using
100 MCMC draws from each.  This resulted in a $p$-value of 0 indicating
that at least one of the densities is different from the others.  To
explore this a bit more, Figure \ref{threeway}  displays the densities and
Table \ref{Severity1} the predicted osteonecrosis grade. Notice from the
table that among all four configurations with High {\tt HDLWk08}, the
highest risk of grade 2 or higher osteonecrosis is when {\tt AlbuminWk07}
and {\tt DexClWk07} are both Low. On the other hand, when {\tt HDLWk08} is
Low, having {\tt AlbuminWk07} and {\tt DexClWk07} both Low does not have
the highest risk of grade 2 or higher osteonecrosis. Instead, the highest
risk of grade 2 or higher osteonecrosis when {\tt HDLWk08} is Low is when
{\tt AlbuminWk07} is Low but {\tt DexClWk07} is High (0.064). Having both
{\tt AlbuminWk07} and {\tt DexClWk07} Low is the second lowest risk
(0.032).  This is also verified by Figure \ref{threeway} where these three
combinations display humps at latent value locations corresponding to the
specific osteonecrosis grades.

\begin{table}[htp]
\caption{Posterior probabilities corresponding to each severity grade of
osteonecrosis for {\tt AlbuminWk07}, {\tt DexClWk07}, and {\tt HDLWk08}}
\begin{center}
\begin{tabular}{ccc|ccc}
\toprule
\multicolumn{3}{c}{} & \multicolumn{3}{c}{Osteonecrosis Grade} \\ \cmidrule(r){4-6}
AlbuminWk07&DexClWk07&HDLWk08&0&1&$\ge$2\\ \midrule
Low & Low & Low & 		0.149 & 0.819 & 0.032 \\
High & Low & Low & 	0.196 & 0.802 & 0.002 \\
Low & High & Low&		0.309 & 0.627 & 0.064 \\
High & High & Low & 	0.428 & 0.560 & 0.012 \\
Low & Low & High &		0.114 & 0.749 & 0.137 \\
High & Low & High &	 	0.327 & 0.662 & 0.011 \\
Low & High & High & 	0.341 & 0.593 & 0.066 \\
High & High & High & 	0.441 & 0.535 & 0.024 \\
\bottomrule
\end{tabular}
\end{center}
\label{Severity1}
\end{table}%

\subsection{Multivariate Similarity Function}

\begin{table}[ht]
\caption{Association rules from the Osteonecrosis data ordinal data model
fit with multivariate similarity functions assigned to the temporally
referenced covariates. Here only association rules that contain either
{\tt DexClwk07},  {\tt DexClwk08}, or {\tt AsparaginaseAntibodyAUC} are
considered} \hspace{-0.25 cm} \resizebox{\textwidth}{!}{
\begin{tabular}{ r@{$\ \Leftrightarrow \ $} l ccccc}
\toprule
  \multicolumn{2}{c}{Association Rule} & $Pr$ & Supp. & Conf.  & $|S|$ & $p$-value \\
\midrule
	DexClWk07 & LowRisk  					& 1.00 & 0.63 & 0.91 & 69.07 & 0.00 \\
  	AsparaginaseAntibodyAUC & AlbuminWk07 	& 0.98 & 0.82 & 0.96 & 52.22 & 0.37 \\
  	DexClWk08 & LowRisk  					& 0.90 & 0.56 & 0.86 & 68.70 & 0.53 \\
  	DexClWk08 & LDLBaseline  				& 0.65 & 0.63 & 0.99 & 15.78 & 0.61 \\
  	DexClWk07 & LDLBaseline  				& 0.61 & 0.62 & 1.00 & 16.06 & 0.00 \\
  	DexClWk07 & AlbuminWk07 				& 0.54 & 0.66 & 0.91 & 22.93 & 0.00 \\
  	AsparaginaseAntibodyAUC & LowRisk 		& 0.51 & 0.82 & 0.96 & 52.03 & 0.12 \\
  	DexClWk08 & AlbuminWk08 				& 0.49 & 0.81 & 0.98 & 13.36 & 0.61 \\
\bottomrule
\end{tabular}
}
\label{aruls4}
\end{table}

In this section we show how the RAID procedure can
accommodate the temporal structure that exists among the covariates
through the PPMx prior. This is done by considering the following
multivariate type similarity:
\begin{align}\label{MVSimilarityFunction}
g(\bm{X}^{\star}_j|\bm{\eta}) = \prod_{\ell = 1}^{p_q} g(\bm{X}^{\star}_{j \ell}|\bm{\eta})
\prod_{\ell=p_q+1}^{p_q+p_c}g(\bm{X}^{\star}_{j \ell}|\bm{\eta})\prod_{\ell=p_q+p_c+1}^{p_q+p_c+p_t}g(\bm{X}^{\star}_{j \ell}|\bm{\eta}),
\end{align}
where  $p_t$ denotes the number of temporally measured variables, and
individual covariate vectors are ordered such that qualitative variables
appear first, categorical second, and time-dependent last.  For temporal
covariates $q(\cdot | \bm{\xi}^{\star})$ and  $q(\bm{\xi}^{\star})$
correspond to a multivariate Gaussian density and a multivariate
Gaussian-Inverse-Wishart density respectively, with dimensionality given
by the number of measurements for that particular covariate. It seems
plausible that this similarity construction will be able to more
accurately measure the compactness of temporal variables as correlations
will be included. Using the same MCMC specifications as in the univariate
similarity case, we fit the ordinal data model with a multivariate PPMx
prior.  The two model fits are fairly compatible although the multivariate
similarity produces a slightly lower log pseudo marginal likelihood (LPML)
(-275.74) relative to the univariate similarity (-255.25) indicating a
slightly worse model fit. In addition, the average number of clusters is
10.3 for the former and 11.2 for the later. Association rule results are
provided in Table \ref{aruls4}. All association rules that appear in Table
\ref{aruls4} also appear in  Table \ref{aruls2} save those that involve
{\tt LDLBaseline}, which is not confirmed when paired with {\tt
DexClWk08}.  Perhaps the biggest difference between  Tables \ref{aruls2}
and \ref{aruls4} is that the association rule between {\tt
AsparaginaseAntibodyAUC} and {\tt AlbuminWk07} is now not confirmed.
Thus, the interaction between these two variables may be changing with
specific values of the covariates that depend on time.  This illustrates
that the exploratory approach depends on the specific model employed.

\subsection{Sensitivity to Prior Specification}

We provide tables containing the list of potential interactions for each
of the 18 prior configurations outlined in Section \ref{priorSensitivity} of the main
document. Recall that for ordinal data model \eqref{latentModel} used to
fit the osteonecrosis data the priors whose sensitivity is being explored
are  $A$, $c(\cdot)$, and $g(\cdot)$. In particular, for the similarity
$g(\cdot)$ and for continuous covariates we vary $k_0$ in
\begin{align*}
g(\bm{X}^{\star}_j|\bm{\eta}) = \prod_{\ell=1}^{p_c}\int \prod_{i \in S_j} N(\bm{X}_{i\ell}|m^{\star},s^{2\star})
N(m^{\star}|m_0, k_0s^{2\star})IG(s^{2\star}| \nu_0, \kappa_0)dm^{\star}ds^{2\star}
\end{align*}
while fixing $m_0 = 0$,  $\nu_0 = 1$, and $\kappa_0 = 2$. The parameter
values employed in the sensitivity analysis are $A \in \{0.1, 1, 10\}$,
$k_0 \in \{0.1, 1, 10\}$ and $c(\cdot) = 1$ or $c(\cdot) = M \times
(|\cdot| - 1)!$  with $M=1$.  The list of interactions detected for each
prior specification are provided in Tables \ref{sensitivity1} -
\ref{sensitivity18}.  As expected there are differences among the 18
tables.  Most notably, when $k_0$ is large (i.e., 10), there are much less
interactions detected. This is to be expected as the connection between
$\bm{X}$ and $\bm{Y}$ is weaker as the values of $\bm{Y}$ are what is
primarily driving cluster formation.   It seems that results are quite
robust to the specification of $c(\cdot)$.

\begin{table}[ht]
\caption{Association rules from the Osteonecrosis data ordinal data model.
Here only association rules that contain either {\tt DexClwk07}, {\tt
DexClwk08}, or {\tt AsparaginaseAntibodyAUC} are considered.  Prior
specification is $A=0.1$, $k_0 = 0.1$, $c(S) = 1$} \hspace{-0.25 cm}
\resizebox{\textwidth}{!}{
\begin{tabular}{ r@{$\ \Leftrightarrow \ $} l ccccc}
\toprule
  \multicolumn{2}{c}{Association Rule} & $Pr$ & Supp. & Conf.  & $|S|$ & $p$-value \\
  \midrule
AsparaginaseAntibodyAUC & Race2 & 0.95 & 0.69 & 1.00 & 8.09 & 0.18 \\
  AsparaginaseAntibodyAUC & HDLWk07 & 0.90 & 0.70 & 1.00 & 8.00 & 0.00 \\
  DexClWk07 & LDLWk07 & 0.70 & 0.81 & 0.99 & 11.13 & 0.00 \\
  DexClWk07 & AlbuminWk07 & 0.69 & 0.70 & 1.00 & 11.86 & 0.00 \\
  AsparaginaseAntibodyAUC & AlbuminWk07 & 0.66 & 0.97 & 0.99 & 18.56 & 0.00 \\
  AsparaginaseAntibodyAUC & TriglyceridesWk08 & 0.66 & 0.98 & 1.00 & 19.32 & 0.38 \\
  DexClWk07 & LDLWk08 & 0.65 & 0.77 & 0.99 & 13.38 & 0.00 \\
  DexClWk07 & AgeAtDiagnosis & 0.62 & 0.77 & 0.98 & 11.73 & 0.00 \\
  DexClWk07 & Race2 & 0.59 & 0.75 & 0.98 & 11.24 & 0.00 \\
  AsparaginaseAntibodyAUC & LowRisk & 0.55 & 1.00 & 1.00 & 19.00 & 0.00 \\
  DexClWk08 & AlbuminWk08 & 0.51 & 0.61 & 1.00 & 7.25 & 0.00 \\
   \bottomrule
\end{tabular}
}
\label{sensitivity1}
\end{table}

\begin{table}[ht]
\caption{Association rules from the Osteonecrosis data ordinal data model.
Here only association rules that contain either {\tt DexClwk07}, {\tt
DexClwk08}, or {\tt AsparaginaseAntibodyAUC} are considered.  Prior
specification is $A=0.1$, $k_0 = 0.1$, $c(S) = M\times(|S|-1)!$ with
$M=1$} \hspace{-0.25 cm} \resizebox{\textwidth}{!}{
\begin{tabular}{ r@{$\ \Leftrightarrow \ $} l ccccc}
\toprule
  \multicolumn{2}{c}{Association Rule} & $Pr$ & Supp. & Conf.  & $|S|$ & $p$-value \\
  \midrule
AsparaginaseAntibodyAUC & Race2 & 0.97 & 0.69 & 1.00 & 7.96 & 0.12 \\
  AsparaginaseAntibodyAUC & HDLWk07 & 0.92 & 0.69 & 1.00 & 7.89 & 0.00 \\
  DexClWk07 & LDLWk07 & 0.83 & 0.81 & 1.00 & 10.89 & 0.00 \\
  AsparaginaseAntibodyAUC & AlbuminWk07 & 0.64 & 0.98 & 1.00 & 19.12 & 0.01 \\
  AsparaginaseAntibodyAUC & TriglyceridesWk08 & 0.63 & 0.98 & 1.00 & 19.56 & 0.37 \\
  DexClWk07 & AlbuminWk07 & 0.62 & 0.72 & 1.00 & 11.63 & 0.00 \\
  DexClWk07 & AgeAtDiagnosis & 0.61 & 0.77 & 0.99 & 10.56 & 0.00 \\
  DexClWk07 & LDLWk08 & 0.60 & 0.75 & 0.99 & 12.80 & 0.00 \\
  DexClWk07 & Race2 & 0.52 & 0.76 & 0.99 & 10.31 & 0.00 \\
   \bottomrule
\end{tabular}
}
\end{table}

\begin{table}[ht]
\caption{Association rules from the Osteonecrosis data ordinal data model.
Here only association rules that contain either {\tt DexClwk07}, {\tt
DexClwk08}, or {\tt AsparaginaseAntibodyAUC} are considered.  Prior
specification is $A=0.1$, $k_0 = 1$, $c(S) = 1$} \hspace{-0.25 cm}
\resizebox{\textwidth}{!}{
\begin{tabular}{ r@{$\ \Leftrightarrow \ $} l ccccc}
\toprule
  \multicolumn{2}{c}{Association Rule} & $Pr$ & Supp. & Conf.  & $|S|$ & $p$-value \\
  \midrule
DexClWk07 & LowRisk & 0.95 & 0.61 & 0.95 & 25.73 & 0.00 \\
  AsparaginaseAntibodyAUC & LowRisk & 0.93 & 0.84 & 1.00 & 27.34 & 0.00 \\
  DexClWk07 & AlbuminWk07 & 0.92 & 0.61 & 0.91 & 21.96 & 0.00 \\
  AsparaginaseAntibodyAUC & AlbuminWk07 & 0.84 & 0.80 & 0.93 & 25.48 & 0.00 \\
  DexClWk07 & LDLWk08 & 0.84 & 0.71 & 0.98 & 21.33 & 0.00 \\
  AsparaginaseAntibodyAUC & LDLWk08 & 0.80 & 0.71 & 0.99 & 21.45 & 0.00 \\
  DexClWk07 & Race2 & 0.77 & 0.69 & 0.92 & 15.53 & 0.00 \\
  AsparaginaseAntibodyAUC & TriglyceridesWk08 & 0.68 & 0.91 & 0.99 & 19.21 & 0.00 \\
   \bottomrule
\end{tabular}
}
\end{table}

\begin{table}[ht]
\caption{Association rules from the Osteonecrosis data ordinal data model.
Here only association rules that contain either {\tt DexClwk07}, {\tt
DexClwk08}, or {\tt AsparaginaseAntibodyAUC} are considered.  Prior
specification is $A=0.1$, $k_0 = 1$, $c(S) = M\times(|S|-1)!$ with $M=1$}
\hspace{-0.25 cm} \resizebox{\textwidth}{!}{
\begin{tabular}{ r@{$\ \Leftrightarrow \ $} l ccccc}
\toprule
  \multicolumn{2}{c}{Association Rule} & $Pr$ & Supp. & Conf.  & $|S|$ & $p$-value \\
  \midrule
AsparaginaseAntibodyAUC & AlbuminWk07 & 0.96 & 0.81 & 0.94 & 25.28 & 0.00 \\
  DexClWk07 & AlbuminWk07 & 0.94 & 0.63 & 0.94 & 19.45 & 0.00 \\
  AsparaginaseAntibodyAUC & LowRisk & 0.93 & 0.85 & 0.99 & 27.52 & 0.00 \\
  DexClWk07 & LDLWk08 & 0.80 & 0.72 & 0.96 & 21.65 & 0.00 \\
  DexClWk07 & Race2 & 0.79 & 0.67 & 0.92 & 15.96 & 0.00 \\
  DexClWk07 & LowRisk & 0.77 & 0.62 & 0.94 & 25.83 & 0.00 \\
  AsparaginaseAntibodyAUC & LDLWk08 & 0.68 & 0.70 & 0.99 & 22.25 & 0.00 \\
  DexClWk08 & HDLWk08 & 0.63 & 0.59 & 0.90 & 23.54 & 0.00 \\
  AsparaginaseAntibodyAUC & TriglyceridesWk08 & 0.58 & 0.90 & 0.99 & 19.28 & 0.00 \\
  AsparaginaseAntibodyAUC & HDLWk07 & 0.56 & 0.76 & 0.97 & 13.05 & 0.00 \\
   \bottomrule
\end{tabular}
}
\end{table}

\begin{table}[ht]
\caption{Association rules from the Osteonecrosis data ordinal data model.
Here only association rules that contain either {\tt DexClwk07}, {\tt
DexClwk08}, or {\tt AsparaginaseAntibodyAUC} are considered. Prior
specification is $A=0.1$, $k_0 = 10$, $c(S) = 1$} \hspace{-0.25 cm}
\resizebox{\textwidth}{!}{
\begin{tabular}{ r@{$\ \Leftrightarrow \ $} l ccccc}
\toprule
  \multicolumn{2}{c}{Association Rule} & $Pr$ & Supp. & Conf.  & $|S|$ & $p$-value \\
  \midrule
DexClWk07 & AlbuminWk07 & 1.00 & 0.60 & 0.89 & 63.87 & 0.00 \\
  DexClWk07 & LowRisk & 1.00 & 0.64 & 0.94 & 67.96 & 0.00 \\
  AsparaginaseAntibodyAUC & AlbuminWk07 & 0.98 & 0.58 & 0.83 & 49.19 & 0.00 \\
  DexClWk08 & LowRisk & 0.97 & 0.75 & 0.96 & 17.44 & 0.00 \\
  DexClWk08 & CortisolWk07 & 0.95 & 0.53 & 1.00 & 10.73 & 0.01 \\
  AsparaginaseAntibodyAUC & LowRisk & 0.89 & 0.58 & 0.87 & 61.00 & 0.00 \\
  AsparaginaseAntibodyAUC & TriglyceridesWk08 & 0.74 & 0.57 & 0.82 & 48.42 & 0.00 \\
  DexClWk08 & AlbuminWk08 & 0.55 & 0.70 & 0.89 & 16.16 & 0.00 \\
   \bottomrule
\end{tabular}
}
\end{table}

\begin{table}[ht]
\caption{Association rules from the Osteonecrosis data ordinal data model.
Here only association rules that contain either {\tt DexClwk07}, {\tt
DexClwk08}, or {\tt AsparaginaseAntibodyAUC} are considered.  Prior
specification is $A=0.1$, $k_0 = 10$, $c(S) = M\times(|S|-1)!$ with $M=1$}
\hspace{-0.25 cm} \resizebox{\textwidth}{!}{
\begin{tabular}{ r@{$\ \Leftrightarrow \ $} l ccccc}
\toprule
  \multicolumn{2}{c}{Association Rule} & $Pr$ & Supp. & Conf.  & $|S|$ & $p$-value \\
  \midrule
DexClWk07 & AlbuminWk07 & 1.00 & 0.60 & 0.89 & 64.14 & 0.00 \\
  DexClWk07 & LowRisk & 1.00 & 0.64 & 0.94 & 68.34 & 0.00 \\
  AsparaginaseAntibodyAUC & AlbuminWk07 & 0.98 & 0.58 & 0.83 & 49.17 & 0.00 \\
  DexClWk08 & LowRisk & 0.98 & 0.75 & 0.96 & 17.36 & 0.00 \\
  AsparaginaseAntibodyAUC & LowRisk & 0.91 & 0.58 & 0.87 & 61.12 & 0.00 \\
  DexClWk08 & CortisolWk07 & 0.82 & 0.53 & 1.00 & 10.55 & 0.00 \\
  AsparaginaseAntibodyAUC & TriglyceridesWk08 & 0.70 & 0.57 & 0.81 & 48.49 & 0.00 \\
  DexClWk08 & AlbuminWk08 & 0.68 & 0.70 & 0.89 & 16.32 & 0.00 \\
   \bottomrule
\end{tabular}
}
\end{table}

\begin{table}[ht]
\caption{Association rules from the Osteonecrosis data ordinal data model.
Here only association rules that contain either {\tt DexClwk07}, {\tt
DexClwk08}, or {\tt AsparaginaseAntibodyAUC} are considered.  Prior
specification is $A=1$, $k_0 = 0.1$, $c(S) = 1$} \hspace{-0.25 cm}
\resizebox{\textwidth}{!}{
\begin{tabular}{ r@{$\ \Leftrightarrow \ $} l ccccc}
\toprule
  \multicolumn{2}{c}{Association Rule} & $Pr$ & Supp. & Conf.  & $|S|$ & $p$-value \\
  \midrule
AsparaginaseAntibodyAUC & AlbuminWk07 & 0.85 & 0.96 & 0.99 & 21.79 & 0.21 \\
  DexClWk07 & Race2 & 0.71 & 0.76 & 0.97 & 12.77 & 0.00 \\
  DexClWk07 & LDLWk07 & 0.67 & 0.83 & 1.00 & 10.96 & 0.02 \\
  AsparaginaseAntibodyAUC & HDLWk07 & 0.60 & 0.73 & 1.00 & 8.42 & 0.00 \\
  AsparaginaseAntibodyAUC & Race2 & 0.59 & 0.71 & 0.98 & 8.63 & 0.20 \\
  DexClWk07 & LowRisk & 0.58 & 0.89 & 1.00 & 15.67 & 0.01 \\
  AsparaginaseAntibodyAUC & LowRisk & 0.56 & 1.00 & 1.00 & 23.16 & 0.08 \\
  AsparaginaseAntibodyAUC & TriglyceridesWk08 & 0.55 & 0.99 & 1.00 & 23.71 & 0.02 \\
  DexClWk07 & AgeAtDiagnosis & 0.55 & 0.78 & 0.97 & 12.47 & 0.04 \\
  DexClWk08 & HDLWk08 & 0.55 & 0.69 & 0.99 & 9.96 & 0.00 \\
  DexClWk07 & LDLWk08 & 0.51 & 0.76 & 0.98 & 14.20 & 0.00 \\
   \bottomrule
\end{tabular}
}
\end{table}

\begin{table}[ht]
\caption{Association rules from the Osteonecrosis data ordinal data model.
Here only association rules that contain either {\tt DexClwk07}, {\tt
DexClwk08}, or {\tt AsparaginaseAntibodyAUC} are considered. Prior
specification is $A=1$, $k_0 = 0.1$, $c(S) = M\times(|S|-1)!$ with $M=1$}
\hspace{-0.25 cm} \resizebox{\textwidth}{!}{
\begin{tabular}{ r@{$\ \Leftrightarrow \ $} l ccccc}
\toprule
  \multicolumn{2}{c}{Association Rule} & $Pr$ & Supp. & Conf.  & $|S|$ & $p$-value \\
  \midrule
AsparaginaseAntibodyAUC & AlbuminWk07 & 0.81 & 0.97 & 0.99 & 22.20 & 0.01 \\
  DexClWk07 & Race2 & 0.76 & 0.74 & 0.97 & 12.67 & 0.00 \\
  DexClWk07 & LDLWk07 & 0.75 & 0.84 & 1.00 & 11.20 & 0.00 \\
  DexClWk07 & LDLWk08 & 0.71 & 0.76 & 0.99 & 14.00 & 0.00 \\
  AsparaginaseAntibodyAUC & Race2 & 0.70 & 0.72 & 0.99 & 8.64 & 0.28 \\
  AsparaginaseAntibodyAUC & HDLWk07 & 0.64 & 0.73 & 1.00 & 8.38 & 0.00 \\
  AsparaginaseAntibodyAUC & LowRisk & 0.63 & 1.00 & 1.00 & 23.32 & 0.03 \\
  DexClWk07 & LowRisk & 0.61 & 0.89 & 0.99 & 16.11 & 0.00 \\
  AsparaginaseAntibodyAUC & TriglyceridesWk08 & 0.59 & 0.99 & 1.00 & 23.86 & 0.05 \\
  DexClWk07 & AgeAtDiagnosis & 0.55 & 0.80 & 0.98 & 12.78 & 0.01 \\
  DexClWk08 & HDLWk08 & 0.52 & 0.70 & 0.99 & 10.38 & 0.00 \\
   \bottomrule
\end{tabular}
}
\end{table}

\begin{table}[ht]
\caption{Association rules from the Osteonecrosis data ordinal data model.
Here only association rules that contain either {\tt DexClwk07}, {\tt
DexClwk08}, or {\tt AsparaginaseAntibodyAUC} are considered. Prior
specification is $A=1$, $k_0 = 1$, $c(S) = 1$} \hspace{-0.25 cm}
\resizebox{\textwidth}{!}{
\begin{tabular}{ r@{$\ \Leftrightarrow \ $} l ccccc}
\toprule
  \multicolumn{2}{c}{Association Rule} & $Pr$ & Supp. & Conf.  & $|S|$ & $p$-value \\
  \midrule
DexClWk07 & LowRisk & 0.96 & 0.58 & 0.90 & 27.27 & 0.00 \\
  DexClWk07 & Race2 & 0.91 & 0.70 & 0.93 & 17.23 & 0.00 \\
  AsparaginaseAntibodyAUC & AgeAtDiagnosis & 0.88 & 0.85 & 0.99 & 11.82 & 0.00 \\
  DexClWk08 & HDLWk08 & 0.85 & 0.57 & 0.88 & 27.20 & 0.00 \\
  DexClWk07 & LDLWk08 & 0.82 & 0.74 & 0.98 & 20.20 & 0.00 \\
  AsparaginaseAntibodyAUC & TriglyceridesWk08 & 0.79 & 0.90 & 0.98 & 29.20 & 0.00 \\
  DexClWk07 & AlbuminWk07 & 0.75 & 0.62 & 0.92 & 20.11 & 0.00 \\
  AsparaginaseAntibodyAUC & LDLWk08 & 0.68 & 0.71 & 1.00 & 19.87 & 0.00 \\
  AsparaginaseAntibodyAUC & LDLWk07 & 0.59 & 0.71 & 0.96 & 18.14 & 0.00 \\
  AsparaginaseAntibodyAUC & AlbuminWk07 & 0.57 & 0.84 & 0.96 & 25.95 & 0.00 \\
  DexClWk07 & LDLWk07 & 0.57 & 0.66 & 0.98 & 16.07 & 0.00 \\
  DexClWk07 & HDLBaseline & 0.52 & 0.75 & 0.97 & 10.77 & 0.00 \\
  AsparaginaseAntibodyAUC & LowRisk & 0.51 & 0.87 & 0.98 & 27.65 & 0.00 \\
   \bottomrule
\end{tabular}
}
\end{table}

\begin{table}[ht]
\caption{Association rules from the Osteonecrosis data ordinal data model.
Here only association rules that contain either {\tt DexClwk07}, {\tt
DexClwk08}, or {\tt AsparaginaseAntibodyAUC} are considered. Prior
specification is $A=1$, $k_0 = 1$, $c(S) = M\times(|S|-1)!$ with $M=1$}
\hspace{-0.25 cm} \resizebox{\textwidth}{!}{
\begin{tabular}{ r@{$\ \Leftrightarrow \ $} l ccccc}
\toprule
  \multicolumn{2}{c}{Association Rule} & $Pr$ & Supp. & Conf.  & $|S|$ & $p$-value \\
  \midrule
DexClWk07 & LowRisk & 0.95 & 0.58 & 0.90 & 27.69 & 0.00 \\
  DexClWk07 & Race2 & 0.95 & 0.70 & 0.93 & 17.15 & 0.00 \\
  DexClWk07 & LDLWk08 & 0.92 & 0.74 & 0.98 & 20.52 & 0.00 \\
  AsparaginaseAntibodyAUC & AgeAtDiagnosis & 0.86 & 0.86 & 1.00 & 11.86 & 0.00 \\
  DexClWk07 & AlbuminWk07 & 0.84 & 0.63 & 0.91 & 20.38 & 0.00 \\
  AsparaginaseAntibodyAUC & LowRisk & 0.78 & 0.86 & 0.96 & 31.00 & 0.00 \\
  DexClWk08 & HDLWk08 & 0.78 & 0.57 & 0.88 & 27.41 & 0.00 \\
  AsparaginaseAntibodyAUC & LDLWk08 & 0.68 & 0.72 & 1.00 & 20.56 & 0.00 \\
  AsparaginaseAntibodyAUC & AlbuminWk07 & 0.68 & 0.82 & 0.95 & 28.01 & 0.00 \\
  AsparaginaseAntibodyAUC & HDLWk08 & 0.56 & 0.74 & 0.93 & 11.79 & 0.00 \\
   \bottomrule
\end{tabular}
}
\end{table}

\begin{table}[ht]
\caption{Association rules from the Osteonecrosis data ordinal data model.
Here only association rules that contain either {\tt DexClwk07}, {\tt
DexClwk08}, or {\tt AsparaginaseAntibodyAUC} are considered. Prior
specification is $A=1$, $k_0 = 10$, $c(S) = 1$} \hspace{-0.25 cm}
\resizebox{\textwidth}{!}{
\begin{tabular}{ r@{$\ \Leftrightarrow \ $} l ccccc}
\toprule
  \multicolumn{2}{c}{Association Rule} & $Pr$ & Supp. & Conf.  & $|S|$ & $p$-value \\
  \midrule
DexClWk07 & AlbuminWk07 & 1.00 & 0.54 & 0.83 & 69.41 & 0.00 \\
  DexClWk07 & LowRisk & 1.00 & 0.59 & 0.91 & 76.47 & 0.00 \\
  DexClWk08 & CortisolWk07 & 0.70 & 0.50 & 0.84 & 10.89 & 0.90 \\
  DexClWk07 & TriglyceridesWk08 & 0.66 & 0.50 & 0.87 & 11.08 & 0.00 \\
  AsparaginaseAntibodyAUC & LowRisk & 0.61 & 0.51 & 0.83 & 65.11 & 0.01 \\
   \bottomrule
\end{tabular}
}
\end{table}

\begin{table}[ht]
\caption{Association rules from the Osteonecrosis data ordinal data model.
Here only association rules that contain either {\tt DexClwk07}, {\tt
DexClwk08}, or {\tt AsparaginaseAntibodyAUC} are considered. Prior
specification is $A=1$, $k_0 = 10$, $c(S) = M\times(|S|-1)!$ with $M=1$}
\hspace{-0.25 cm} \resizebox{\textwidth}{!}{
\begin{tabular}{ r@{$\ \Leftrightarrow \ $} l ccccc}
\toprule
  \multicolumn{2}{c}{Association Rule} & $Pr$ & Supp. & Conf.  & $|S|$ & $p$-value \\
  \midrule
DexClWk07 & AlbuminWk07 & 1.00 & 0.53 & 0.82 & 69.48 & 0.00 \\
  DexClWk07 & LowRisk & 1.00 & 0.59 & 0.91 & 76.78 & 0.00 \\
  DexClWk07 & TriglyceridesWk08 & 0.83 & 0.50 & 0.86 & 11.30 & 0.00 \\
  DexClWk08 & CortisolWk07 & 0.66 & 0.50 & 0.85 & 11.11 & 1.00 \\
  AsparaginaseAntibodyAUC & LowRisk & 0.63 & 0.50 & 0.83 & 65.16 & 0.12 \\
  DexClWk07 & TriglyceridesWk07 & 0.51 & 0.66 & 0.87 & 52.43 & 0.00 \\
   \bottomrule
\end{tabular}
}
\end{table}

\begin{table}[ht]
\caption{Association rules from the Osteonecrosis data ordinal data model.
Here only association rules that contain either {\tt DexClwk07}, {\tt
DexClwk08}, or {\tt AsparaginaseAntibodyAUC} are considered. Prior
specification is $A=10$, $k_0 = 0.1$, $c(S) = 1$} \hspace{-0.25 cm}
\resizebox{\textwidth}{!}{
\begin{tabular}{ r@{$\ \Leftrightarrow \ $} l ccccc}
\toprule
  \multicolumn{2}{c}{Association Rule} & $Pr$ & Supp. & Conf.  & $|S|$ & $p$-value \\
  \midrule
DexClWk07 & Race2 & 0.82 & 0.76 & 0.97 & 13.24 & 0.00 \\
  DexClWk08 & HDLWk08 & 0.74 & 0.72 & 0.99 & 13.64 & 0.00 \\
  AsparaginaseAntibodyAUC & HDLWk07 & 0.72 & 0.74 & 1.00 & 8.58 & 0.00 \\
  AsparaginaseAntibodyAUC & AlbuminWk07 & 0.71 & 0.93 & 0.99 & 21.04 & 0.12 \\
  AsparaginaseAntibodyAUC & Race2 & 0.67 & 0.72 & 0.98 & 8.57 & 0.46 \\
  DexClWk07 & AgeAtDiagnosis & 0.62 & 0.78 & 0.96 & 14.45 & 0.00 \\
  DexClWk07 & LDLWk08 & 0.60 & 0.78 & 0.98 & 15.42 & 0.00 \\
  DexClWk08 & AlbuminWk07 & 0.55 & 0.70 & 0.96 & 13.36 & 0.00 \\
  AsparaginaseAntibodyAUC & LowRisk & 0.53 & 1.00 & 1.00 & 24.57 & 0.17 \\
  AsparaginaseAntibodyAUC & LDLWk07 & 0.52 & 0.76 & 0.99 & 11.63 & 0.15 \\
  DexClWk07 & LowRisk & 0.51 & 0.88 & 0.99 & 16.78 & 0.00 \\
   \bottomrule
\end{tabular}
}
\end{table}

\begin{table}[ht]
\caption{Association rules from the Osteonecrosis data ordinal data model.
Here only association rules that contain either {\tt DexClwk07}, {\tt
DexClwk08}, or {\tt AsparaginaseAntibodyAUC} are considered. Prior
specification is $A=10$, $k_0 = 0.1$, $c(S) = M\times(|S|-1)!$ with $M=1$}
\hspace{-0.25 cm} \resizebox{\textwidth}{!}{
\begin{tabular}{ r@{$\ \Leftrightarrow \ $} l ccccc}
\toprule
  \multicolumn{2}{c}{Association Rule} & $Pr$ & Supp. & Conf.  & $|S|$ & $p$-value \\
  \midrule
DexClWk08 & HDLWk08 & 0.78 & 0.70 & 0.98 & 14.17 & 0.00 \\
  DexClWk07 & Race2 & 0.77 & 0.77 & 0.97 & 12.96 & 0.00 \\
  AsparaginaseAntibodyAUC & HDLWk07 & 0.74 & 0.73 & 1.00 & 8.41 & 0.00 \\
  AsparaginaseAntibodyAUC & AlbuminWk07 & 0.73 & 0.92 & 0.98 & 20.04 & 0.09 \\
  AsparaginaseAntibodyAUC & Race2 & 0.67 & 0.71 & 0.98 & 8.18 & 0.55 \\
  DexClWk07 & LowRisk & 0.58 & 0.88 & 1.00 & 16.79 & 0.00 \\
  AsparaginaseAntibodyAUC & LowRisk & 0.57 & 0.90 & 1.00 & 18.88 & 0.02 \\
  DexClWk07 & AgeAtDiagnosis & 0.54 & 0.78 & 0.96 & 14.52 & 0.02 \\
  DexClWk08 & AlbuminWk07 & 0.54 & 0.71 & 0.95 & 14.15 & 0.00 \\
  DexClWk07 & LDLWk08 & 0.53 & 0.78 & 0.98 & 15.09 & 0.00 \\
  DexClWk07 & AlbuminWk07 & 0.52 & 0.69 & 0.99 & 12.58 & 0.00 \\
   \bottomrule
\end{tabular}
}
\end{table}

\begin{table}[ht]
\caption{Association rules from the Osteonecrosis data ordinal data model.
Here only association rules that contain either {\tt DexClwk07}, {\tt
DexClwk08}, or {\tt AsparaginaseAntibodyAUC} are considered. Prior
specification is $A=10$, $k_0 = 1$, $c(S) = 1$} \hspace{-0.25 cm}
\resizebox{\textwidth}{!}{
\begin{tabular}{ r@{$\ \Leftrightarrow \ $} l ccccc}
\toprule
  \multicolumn{2}{c}{Association Rule} & $Pr$ & Supp. & Conf.  & $|S|$ & $p$-value \\
  \midrule
DexClWk07 & Race2 & 0.99 & 0.70 & 0.93 & 16.98 & 0.00 \\
  DexClWk07 & LowRisk & 0.97 & 0.56 & 0.90 & 27.46 & 0.00 \\
  DexClWk07 & LDLWk08 & 0.88 & 0.75 & 0.98 & 20.91 & 0.00 \\
  DexClWk08 & HDLWk08 & 0.85 & 0.55 & 0.87 & 27.46 & 0.00 \\
  AsparaginaseAntibodyAUC & AgeAtDiagnosis & 0.83 & 0.86 & 1.00 & 12.11 & 0.00 \\
  AsparaginaseAntibodyAUC & LowRisk & 0.78 & 0.86 & 0.96 & 32.12 & 0.00 \\
  DexClWk07 & AlbuminWk07 & 0.75 & 0.62 & 0.91 & 20.85 & 0.00 \\
  AsparaginaseAntibodyAUC & LDLWk08 & 0.71 & 0.72 & 0.99 & 20.23 & 0.00 \\
  AsparaginaseAntibodyAUC & AlbuminWk07 & 0.67 & 0.82 & 0.95 & 30.07 & 0.00 \\
  DexClWk07 & CortisolBaseline & 0.55 & 0.89 & 0.98 & 13.04 & 0.00 \\
    \bottomrule
\end{tabular}
}
\end{table}

\begin{table}[ht]
\caption{Association rules from the Osteonecrosis data ordinal data model.
Here only association rules that contain either {\tt DexClwk07}, {\tt
DexClwk08}, or {\tt AsparaginaseAntibodyAUC} are considered. Prior
specification is $A=10$, $k_0 = 1$, $c(S) = M\times(|S|-1)!$ with $M=1$}
\hspace{-0.25 cm} \resizebox{\textwidth}{!}{
\begin{tabular}{ r@{$\ \Leftrightarrow \ $} l ccccc}
\toprule
  \multicolumn{2}{c}{Association Rule} & $Pr$ & Supp. & Conf.  & $|S|$ & $p$-value \\
  \midrule
DexClWk07 & LowRisk & 0.97 & 0.57 & 0.89 & 27.87 & 0.00 \\
  DexClWk07 & LDLWk08 & 0.91 & 0.75 & 0.98 & 20.85 & 0.00 \\
  DexClWk07 & Race2 & 0.91 & 0.70 & 0.93 & 17.23 & 0.00 \\
  AsparaginaseAntibodyAUC & AgeAtDiagnosis & 0.90 & 0.85 & 1.00 & 11.81 & 0.00 \\
  DexClWk08 & HDLWk08 & 0.85 & 0.56 & 0.87 & 27.54 & 0.00 \\
  AsparaginaseAntibodyAUC & LowRisk & 0.81 & 0.82 & 0.95 & 28.72 & 0.00 \\
  DexClWk07 & AlbuminWk07 & 0.80 & 0.63 & 0.92 & 19.44 & 0.00 \\
  AsparaginaseAntibodyAUC & AlbuminWk07 & 0.79 & 0.80 & 0.95 & 27.65 & 0.00 \\
  AsparaginaseAntibodyAUC & LDLWk08 & 0.63 & 0.72 & 1.00 & 20.70 & 0.00 \\
  DexClWk07 & CortisolBaseline & 0.62 & 0.88 & 0.96 & 13.42 & 0.00 \\
  DexClWk07 & HDLBaseline & 0.53 & 0.74 & 0.97 & 12.42 & 0.00 \\
   \bottomrule
\end{tabular}
}
\end{table}

\begin{table}[ht]
\caption{Association rules from the Osteonecrosis data ordinal data model.
Here only association rules that contain either {\tt DexClwk07}, {\tt
DexClwk08}, or {\tt AsparaginaseAntibodyAUC} are considered. Prior
specification is $A=10$, $k_0 = 10$, $c(S) = 1$} \hspace{-0.25 cm}
\resizebox{\textwidth}{!}{
\begin{tabular}{ r@{$\ \Leftrightarrow \ $} l ccccc}
\toprule
  \multicolumn{2}{c}{Association Rule} & $Pr$ & Supp. & Conf.  & $|S|$ & $p$-value \\
  \midrule
DexClWk07 & AlbuminWk07 & 1.00 & 0.54 & 0.83 & 69.71 & 0.00 \\
  DexClWk07 & LowRisk & 1.00 & 0.59 & 0.91 & 76.79 & 0.00 \\
  DexClWk07 & TriglyceridesWk08 & 0.81 & 0.50 & 0.87 & 11.25 & 0.00 \\
  AsparaginaseAntibodyAUC & LowRisk & 0.66 & 0.51 & 0.83 & 65.47 & 0.03 \\
  DexClWk08 & CortisolWk07 & 0.66 & 0.51 & 0.85 & 11.15 & 0.96 \\
   \bottomrule
\end{tabular}
}
\end{table}

\begin{table}[ht]
\caption{Association rules from the Osteonecrosis data ordinal data model.
Here only association rules that contain either {\tt DexClwk07}, {\tt
DexClwk08}, or {\tt AsparaginaseAntibodyAUC} are considered. Prior
specification is $A=10$, $k_0 = 10$, $c(S) = M\times(|S|-1)!$ with $M=1$}
\hspace{-0.25 cm} \resizebox{\textwidth}{!}{
\begin{tabular}{ r@{$\ \Leftrightarrow \ $} l ccccc}
\toprule
  \multicolumn{2}{c}{Association Rule} & $Pr$ & Supp. & Conf.  & $|S|$ & $p$-value \\
  \midrule
DexClWk07 & AlbuminWk07 & 1.00 & 0.53 & 0.82 & 70.08 & 0.00 \\
  DexClWk07 & LowRisk & 1.00 & 0.59 & 0.91 & 77.47 & 0.00 \\
  DexClWk07 & TriglyceridesWk08 & 0.81 & 0.50 & 0.87 & 11.07 & 0.00 \\
  DexClWk08 & CortisolWk07 & 0.78 & 0.51 & 0.85 & 11.13 & 0.99 \\
  AsparaginaseAntibodyAUC & LowRisk & 0.64 & 0.50 & 0.83 & 65.78 & 0.03 \\
  DexClWk07 & TriglyceridesWk07 & 0.57 & 0.66 & 0.87 & 52.25 & 0.00 \\
   \bottomrule
\end{tabular}
}
\label{sensitivity18}
\end{table}

\bibliographystyle{../asa}
\bibliography{../reference}